\documentclass[lettersize,journal]{IEEEtran}

\usepackage{amsmath,amssymb}
\usepackage{fullpage,setspace}
\usepackage{graphicx,epstopdf}
\usepackage{caption,subcaption}

\usepackage{cancel}

\usepackage{color,xcolor}

\begin{document}


\title{Modeling measurements for quantitative imaging subsurface
  targets
  \thanks{This work was supported by the AFOSR Grant
    FA9550-24-1-0191. A.~D.~Kim also acknowledges support by NSF Grant
    DMS-1840265.}}

\author{Arnold D.~Kim and Chrysoula Tsogka
  \thanks{Department of Applied Mathematics, University of California,
  Merced, 5200 North Lake Road, Merced, CA 95343, USA (e-mail:
  adkim@ucmerced.edu and ctsogka@ucmerced.edu)}}

\maketitle

\begin{abstract}
  We study ground-penetrating synthetic aperture radar measurements of
  scattering by targets located below a rough air-soil interface.  By
  considering the inherent space/angle limitations of this imaging
  modality, we introduce a simplified model for measurements. This
  model assumes (i) first-order interactions between the target and
  the air-soil interface, (ii) scattering by the target below a flat
  air-soil interface, and (iii) a point target model. Using the method
  of fundamental solutions to simulate two-dimensional simulations of
  scalar waves for the direct scattering problem, we systematically
  study each of these data modeling assumptions. To test and validate
  these assumptions, we apply principal component analysis to
  approximately remove ground bounce signals from measurements and
  then apply Kirchhoff migration on that processed data to produce
  images.  We show that images using this modeled data are nearly
  identical to those that use simulated measurements from the full
  direct scattering problem. In that way, we show that this model
  contains the essential information contained in
  measurements. Consequently, it provides a theoretical framework for
  understanding how inherent space/angle limitations affect subsurface
  imaging systems.
\end{abstract}


{\em This work has been submitted to the IEEE for possible
publication. Copyright may be transferred without notice, after
which this version may no longer be accessible.} 

\section{Introduction}

\IEEEPARstart{L}{andmines} pose a serious issue for humanity as they
remain dangerous for decades after their deployment, killing and
injuring civilians and rendering the land inaccessible. The economic
burden of landmines is huge, especially for developing countries
\cite{bier}. It is therefore crucially important to develop safe and
robust methods for detection of buried landmines, so that they may be
subsequently removed.

To locate landmines safely, non-touch-based sensors such as ground
penetrating radar (GPR) or metal detectors are preferable \cite{rand}.
GPR can detect both metal and plastic landmines and its weight is
light so that it can be mounted on a unmanned aerial vehicle (UAV)
allowing for safe inspection flying above the air-soil interface as
proposed in \cite{garcia-fernandez:2018, garcia2019bistatic,
  baur:2024}. It is this UAV configuration that we consider here.
Specifically, we consider a sequence of measurements taken a various
locations along a prescribed flight path making up a synthetic
aperture, which we call ground-penetrating synthetic aperture radar
(GP-SAR).

Using UAVs to acquire GP-SAR measurements allows for flight paths at
relatively low elevations above air-soil interfaces. These
low-elevation flight paths help to mitigate environmental factors that
would introduce strong position, navigation, and timing
noises~\cite{garcia-fernandez:2018}. Assuming that buried landmines
are spaced apart relatively far from one another (meters) compared to
the central wavelength used in the imaging system (centimeters), we
consider using relatively short flight paths for relatively small,
site-specific imaging regions. Shorter flight paths imply fewer
measurements which, in turn, require less data storage and faster
processing. These factors may yield efficiency gains through relieving
the communication burden by the UAV, for example.  The UAV may
repeatedly apply this rapid site-specific method to cover larger
regions.

There are several challenges in this problem.  For example, the
air-soil interface is generally not flat and is unknown. The
electromagnetic properties of soil vary widely. They may include
dispersion and absoprtion, for example, and are generally unknown.
Potential subsurface targets also vary widely in shape, size, and
electromagnetic properties.  Measurements of signals scattered by
subsurface targets involve complex scattering by the rough air-soil
interface, the subsurface targets, and the multiple interactions
between them. Various factors of uncertainty always lead to noisy
measurements.

In this work, we seek to determine a model for GP-SAR measurements to
be used for subsurface imaging. We aim to have this model strike a
balance between the complexities of the problem discussed above and
the limitations inherent in this imaging modality. Additionally, we
want this model to be easily extensible to include more
sophistication, as needed.

We determine a GP-SAR measurement model by systematically testing and
validating a sequence of assumptions within the context of a specific
imaging method. This imaging method first uses principal component
analysis (PCA) to approximately remove ground bounce signals, and then
applies Kirchhoff migration (KM) on that processed data to produce an
image that identifies and locates subsurface targets. The
electromagnetic properties of soil need not be known to apply PCA for
ground bounce removal. However, we assume here that the ground
permittivity is known for KM.

We start with a direct scattering problem in which a subsurface target
is situated below a rough air-soil interface. We assume that the size
of the target and the depth below the air-soil interface where it is
situated are on the order of centimeters and therefore comparable to
the central wavelength.  We limit our attention here to
two-dimensional scalar wave propagation where the soil and target are
characterized by their respective relative dielectric
constants. Despite these simplifications, this problem includes strong
scattering by the rough air-soil interface, and multiple scattering
interactions between the target and this interface. These interactions
are the crucially important physical factors affecting measured
signals in this problem which make this imaging problem especially
challenging. Additionally, we include additive measurement noise in
measurements.

The GP-SAR imaging problem is a severely limited aperture imaging
problem.  Because of the inherent space/angle limitations in GP-SAR
measurements, we find that the imaging method we use here produces an
image with limited spatial information about the target --
essentially, images only show a representative single point. It is
most likely that any imaging method will behave similarly because of
these inherent limitations in measurements. In light of these results,
we consider the following sequence of simplifying assumptions:
\begin{enumerate}

\item[(i)] First-order interactions between the target and air-soil
  interface

\item[(ii)] Flat air-soil interface approximation for scattering by
  the target

\item[(iii)] A point target model

\end{enumerate}
In this paper, we systematically test and validate each of these
assumptions in sequence and show that the images resulting from
approximating measurements with them closely match those using the
solution of the full direct scattering problem. From these results, we
are able to propose a simplified model for measurements. This
simplified model immediately provides valuable insight into how the
imaging method we use here works. Moreover, this simplified model
provides a theoretical framework on which one may naturally introduce
extensions for more complex imaging problems.

The remainder of the paper is as follows. In Section
\ref{sec:forward-problem} we discuss the boundary value problem for
the direct scattering problem we study here. In Section \ref{sec:MFS}
we discuss the method of fundamental solutions which we use to solve
this direct scattering problem. We give the details of the imaging
method we use here to evaluate approximations in Section
\ref{sec:inverse-problem}. Our main results are given in Section
\ref{sec:model} where we systematically introduce, test, and validate
the three simplifying assumptions leading to the determination of the
measurement model. We give our conclusions in Section
\ref{sec:conclusions}.

\section{The direct scattering problem}
\label{sec:forward-problem}

\begin{figure}[htb]
  \centering
  \includegraphics[width=0.80\linewidth]{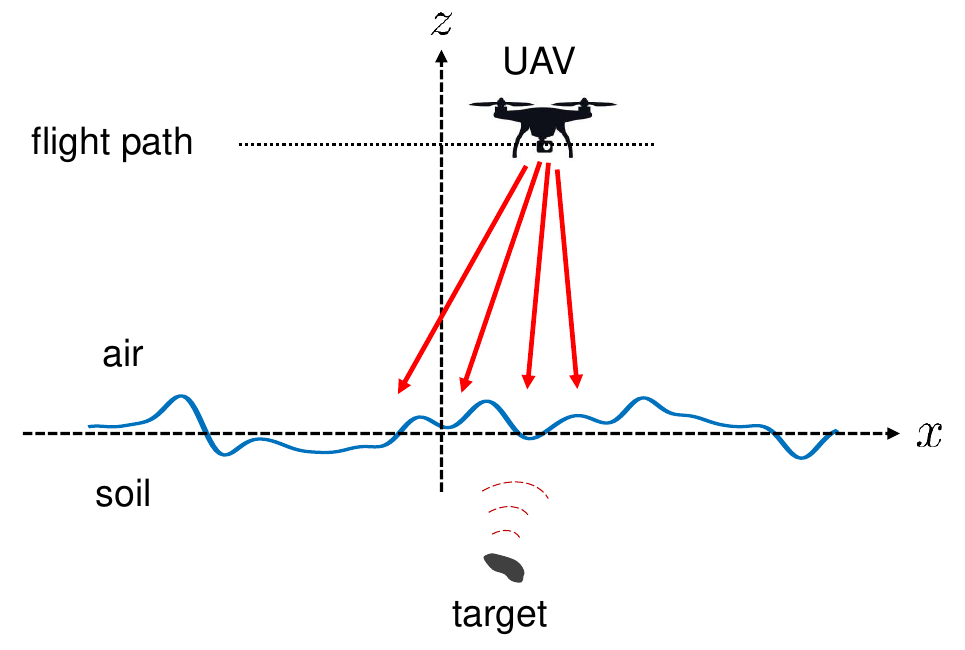}
  \caption{A sketch of the problem in which a UAV emits a signal that
    propagates down onto an air-soil interface. The UAV then takes
    measurements of the response to this source. The objective is to
    reconstruct subsurface targets from these measurements.}
    \label{fig:UAV+SAR}
\end{figure}

A sketch of the physical problem is illustrated in
Fig.~\ref{fig:UAV+SAR}.  As a UAV travels along its flight path, it
emits a multi-frequency signal that propagates down towards the
air-soil interface. Part of this signal is reflected by the air-soil
interface and the other part penetrates into soil. The part of the
signal that has penetrated into the soil is then scattered by a
subsurface target and the part of this scattered signal that transmits
back across the air-soil interface into air is subsequently recorded
by a receiver on the UAV. In this work, we assume the start-stop
approximation which neglects the motion of the UAV between emission
and reception of each measurement along the flight path. Doing so
allows us to focus our discussion on the problem for a fixed source
location at a fixed frequency since measurements consist of repeated
solution of this problem for different source locations and
frequencies.

We describe the boundary value problem governing scalar wave
scattering by a target located below an air-soil interface.  Let
$z = h(x,y)$ denote this air-soil interface. Below this air-soil
interface is a target occupying region $\Omega$ with boundary
$\partial\Omega$. We define the following three regions.
\begin{itemize}

\item Region 0 $= \{ z > h(x,y) \}$ (air)

\item Region 1 $= \{ z < h(x,y) \backslash \Omega \cup \partial\Omega
  \}$ (soil exterior to the target)

\item Region 2 $= \Omega$ (interior to the target)

\end{itemize}
We consider the medium in each of these three regions to be uniform so
that each is characterized solely by their respective wavenumbers:
$k_{0} = \omega/c$ in Region 0, $k_{1} = k_{0} \sqrt{\epsilon_{r1}}$
in Region 1, and $k_{2} = k_{0} \sqrt{\epsilon_{r2}}$ in Region
2. Here, $\omega$ is the circular frequency, $c$ is the wavespeed, and
$\epsilon_{r1}$ and $\epsilon_{r2}$ are the relative dielectric
constants for the soil and the target, respectively.

Let $u_{j}$ denote the scattered field in Region $j$. These fields
satisfy
\begin{subequations}
  \begin{alignat}{2}
    \left( \nabla^{2} + k_{0}^{2} \right) u_{0}
    &= 0, \quad &\text{in Region 0}, \label{eq:PDE0}\\
    \left( \nabla^{2} + k_{1}^{2} \right) u_{1}
    &= 0, \quad &\text{in Region 1}, \label{eq:PDE1}\\
    \left( \nabla^{2} + k_{2}^{2} \right) u_{2}
    &= 0, \quad &\text{in Region 2}, \label{eq:PDE2}
  \end{alignat}
\end{subequations}
with
$\nabla^{2} = \partial_{x}^{2} + \partial_{y}^{2} + \partial_{z}^{2}$
denoting the Laplacian. The incident field in Region 0 is the
free-space Green's function (in air) centered at position
$\mathbf{r}_{\text{src}}$, which we denote by
$u^{\text{inc}} = G_{0}(\mathbf{r} - \mathbf{r}_{\text{src}})$. We
supplement these equations above by boundary conditions. On
$z = h(x,y)$ (the air-soil interface), we require
\begin{equation}
  G_{0} + u_{0} = u_{1} \quad
  \text{and} \quad
  \partial_{n} G_{0} + \partial_{n} u_{0} = \epsilon_{r1}^{-1}
  \partial_{n} u_{1},
  \label{eq:InterfaceBCs}
\end{equation}
with $\partial_{n}$ denoting the normal derivative for the air-soil
interface. On $\partial\Omega$ (the target boundary), we require
\begin{equation}
  u_{1} = u_{2} \quad
  \text{and} \quad
  \epsilon_{r1}^{-1} \partial_{\nu} u_{1} = \epsilon_{r2}^{-1}
  \partial_{\nu} u_{2},
  \label{eq:TargetBCs}
\end{equation}
with $\partial_{\nu}$ denoting the normal derivative for the target
boundary.  Additionally, we require appropriate outgoing conditions
for all of these scattered fields.

\section{The method of fundamental solutions}
\label{sec:MFS}

To solve the system consisting of \eqref{eq:PDE0}, \eqref{eq:PDE1},
and \eqref{eq:PDE2} subject to \eqref{eq:InterfaceBCs},
\eqref{eq:TargetBCs}, and outgoing coniditions, we use the method of
fundamental solutions (MFS). This method was introduced by Mathon and
Johnston~\cite{mathon1977approximate}. It provides an accurate and
efficient computational method for solving the full scattering
problem. The text by Wriedt {\it et al}~\cite{wriedt} provides an
overview of this method and its applications to various problems.  A
recent review paper by Cheng and Hong~\cite{CHENG2020118} focuses on
solvability, uniqueness, convergence and stability of the MFS.

This method most closely resembles integral equation methods, but is
simpler to understand and implement. Just like integral equation
methods, the MFS is a so-called meshless method in that it does not
rely on an underlying mesh over the domain. We have successfully
applied the MFS to multiple scattering problems involving randomly
oriented ellipsoids in an uniform and unbounded
medium~\cite{kim2019intensity}. To our knowledge, the application of
MFS to solve the direct scattering problem here with a rough air-soil
interface has not been done.

The key idea behind the MFS is to introduce approximations for the
scattered fields: $u_{0}$, $u_{1}$, and $u_{2}$, that {\em exactly}
solve \eqref{eq:PDE0}, \eqref{eq:PDE1}, and \eqref{eq:PDE2},
respectively. These approximations are superpositions of finitely-many
fundamental solutions (free-space Green's functions) whose source
positions lie outside of the region. Because their source points lie
outside of the region, the evaluation of these fundamental solutions
within the region exactly solve the differential equation. Moreover,
since each of those fundamental solutions satisfies the appropriate
outgoing conditions, so does their superposition and hence, so do the
MFS approximations.  We determine the relative weighting of each
individual fundamental solution by a collocation method for the
boundary conditions. Thus, the MFS approximation exactly solves the
differential equation and approximately satisfies the boundary
conditions through what is tauntamount to an interpolation
procedure. Because of this, the MFS is able to obtain spectral
accuracy uniformly throughout the domain.

The underlying challenge in using the MFS lies in choosing the
positions for the source points. Typically, they are chosen to attempt
to sample uniformly the boundary/interface but are placed slightly
away from it and outside of the region. There is a natural
relationship between the distance away from the boundary/interface
where source points are placed, the conditioning of the resulting
linear system to determine the weights of the superposition of
fundamental solutions, and the accuracy of the resulting
approximation. When those source points are closer to the
boundary/interface, the resulting linear system is more diagonally
dominant leading to better conditioning, but the approximation is less
accurate. On the other hand, the farther away those source points are
from the boundary/interface, the smoother the approximation is leading
to higher accuracy, but with worse conditioning. There is some theory
on the optimal placement of source points~\cite{barnett2008stability},
but that is limited to two-dimensional problems.

In what follows, we describe our implementation of the MFS for the
direct scattering problem consisting of \eqref{eq:PDE0},
\eqref{eq:PDE1}, and \eqref{eq:PDE2} subject to
\eqref{eq:InterfaceBCs}, \eqref{eq:TargetBCs}, and outgoing
conditions. We limit this discussion to scattering in the
two-dimensional $xz$-plane. Extending this method to three-dimensional
problems is straight-forward in principle, but requires additional
considerations for efficient solution which we do not address here.

\subsection{MFS approximations}

In two dimensions, the air-soil interface is $z = h(x)$. Let
$x_{p} = -L/2 + (p-1)L/P$ for $p = 1, \dots, P$ denote a grid of $P$
points that uniformly samples the interval $-L/2 \le x < L/2$ with
meshwidth $L/P$. We then use the points
$\mathbf{r}^{\text{int}}_{p} = (x_{p}, h(x_{p}))$ for $p = 1, \dots, P$ to
sample the air-soil interface. At $(x_{p}, h(x_{p}))$ we use as the
unit normal to this air-soil interface,
$\hat{\mathbf{n}}_{p} = (-h'(x_{p}),1)/\sqrt{1 +
  (h'(x_{p}))^{2}}$. Note that this choice always points into Region
0.

We assume the target boundary $\partial\Omega$ is a closed curve that
can be parameterized according to $(\xi(t),\zeta(t))$ for
$0 \le t \le 2 \pi$ with
$(\xi(2\pi),\zeta(2\pi)) = (\xi(0),\zeta(0))$. Let
$t_{q} = 2 \pi (q-1) / Q$ for $q = 1, \dots, Q$ denote a grid of $Q$
points that uniformly samples $0 \le t < 2 \pi$ with meshwidth
$2\pi/Q$. We then use the points
$\mathbf{r}_{q}^{\text{tar}} = (\xi(t_{q}), \zeta(t_{q}))$ for
$q = 1, \dots, Q$ to sample the target boundary.  At
$( \xi(t_{q}), \zeta(t_{q}) )$ we use as the unit normal to this
target boundary,
$\hat{\boldsymbol{\nu}}_{q} = ( \zeta'(t_{q}), -\xi'(t_{q}) )/\sqrt{ (
  \xi'(t_{q}) )^{2} + ( \zeta(t_{q}) )^{2} }$. Note that this choice
always points into Region 1.

We introduce two, user-defined parameters: $\delta^{\text{int}} > 0$
and $\delta^{\text{tar}} > 0$. The MFS approximation for scattered
field $u_{0}$ in Region 0 is then
\begin{equation}
  u_{0}(\mathbf{r}) \approx \sum_{p = 1}^{P} G_{0}( \mathbf{r} -
  ( \mathbf{r}^{\text{int}}_{p} - \delta^{\text{int}} \hat{\mathbf{z}}
  ) ) a_{p},
  \label{eq:MFS0}
\end{equation}
with
$G_{0}(\mathbf{r}) = \mathrm{i} \frac{1}{4} H_{0}^{(1)}(k_{0} |
\mathbf{r} |)$ denoting the free-space Green's function with
wavenumber $k_{0}$ and $\mathbf{r}$ denoting any point in Region
0. Note that the source positions:
$\mathbf{r}^{\text{int}}_{p} - \delta^{\text{int}} \hat{\mathbf{z}}$
for $p = 1, \dots, P$ lie outside of Region 0 (inside Region 1), so
\eqref{eq:MFS0} exactly solves \eqref{eq:PDE0} in Region 0. The
expansion coefficients $a_{p}$ for $p = 1, \dots, P$ are to be
determined.

The MFS approximation for scattered field $u_{1}$ in Region 1 is given
by
\begin{multline}
  u_{1}(\mathbf{r}) \approx \sum_{p = 1}^{P} G_{1}( \mathbf{r} -
  ( \mathbf{r}^{\text{int}}_{p} + \delta^{\text{int}} \hat{\mathbf{z}} ) ) b_{p} \\
  + \sum_{q = 1}^{Q} G_{1}( \mathbf{r} - ( \mathbf{r}^{\text{tar}}_{q} -
  \delta^{\text{tar}} \hat{\boldsymbol{\nu}}_{q} ) ) c^{\text{ext}}_{q},
  \label{eq:MFS1}
\end{multline}
with
$G_{1}(\mathbf{r}) = \mathrm{i} \frac{1}{4} H_{0}^{(1)}(k_{1} |
\mathbf{r} |)$ denoting the free-space Green's function with
wavenumber $k_{1}$ and $\mathbf{r}$ denoting any point in Region
1. Note that the source points
$\mathbf{r}^{\text{int}}_{p} + \delta^{\text{int}} \hat{\mathbf{z}}$ for
$p = 1, \dots, P$ and
$\mathbf{r}^{\text{tar}}_{q} - \delta^{\text{tar}} \hat{\boldsymbol{\nu}}_{q}$ for
$q = 1, \dots, Q$ lie outside of Region 1 (inside of Region 0 and
Region 2, respectively), so \eqref{eq:MFS1} exactly solves
\eqref{eq:PDE1} in Region 1. The expansion coefficients $b_{p}$ for
$p = 1, \dots, P$ and $c^{\text{ext}}_{q}$ for $q = 1, \dots, Q$ are
to be determined.

Finally, the MFS approximation for the field interior to the target
$u_{2}$ in Region 2 is
\begin{equation}
  u_{2}(\mathbf{r}) \approx \sum_{q = 1}^{Q} G_{2}( \mathbf{r} -
  ( \mathbf{r}^{\text{tar}}_{q} + \delta^{\text{tar}}
  \hat{\boldsymbol{\nu}}_{q} ) ) c^{\text{int}}_{q},
  \label{eq:MFS2}
\end{equation}
with
$G_{2}(\mathbf{r}) = \mathrm{i} \frac{1}{4} H_{0}^{(1)}(k_{2} |
\mathbf{r} |)$ denoting the free-space Green's function with
wavenumber $k_{2}$ and $\mathbf{r}$ denoting any point in Region
2. Since the source points
$\mathbf{r}^{\text{tar}}_{q} + \delta^{\text{tar}}
\hat{\boldsymbol{\nu}}_{q}$ for $q = 1, \dots, Q$ lie outside of
Region 2 (inside Region 1), \eqref{eq:MFS2} exactly solves
\eqref{eq:PDE2} in Region 2. The expansion coefficients
$c^{\text{int}}_{q}$ for $q = 1, \dots, Q$ are to be determined.

MFS approximations \eqref{eq:MFS0}, \eqref{eq:MFS1}, and
\eqref{eq:MFS2} are all just superpositions of free-space Green's
functions whose source positions lie outside of the regions where they
are evaluated. Additionally, each of these free-space Green's
functions satisfy the correct outgoing conditions. Therefore,
\eqref{eq:MFS0}, \eqref{eq:MFS1}, and \eqref{eq:MFS2} satisfy the
appropriate outgoing conditions in their respective regions.

\subsection{Collocation method for boundary conditions}

In \eqref{eq:MFS0}, \eqref{eq:MFS1}, and \eqref{eq:MFS2}, there are
$2P + 2Q$ undetermined coefficients: $a_{p}$ and $b_{p}$ for
$p = 1, \dots, P$, and $c_{q}^{\text{ext}}$ and $c_{q}^{\text{int}}$
for $q = 1, \dots, Q$.  To determine these undetermined expansion
coefficients, we apply a collocation method for boundary conditions
\eqref{eq:InterfaceBCs} and \eqref{eq:TargetBCs}.

For boundary conditions \eqref{eq:InterfaceBCs} we require that
\begin{multline}
  G_{0}(\mathbf{r}_{p}^{\text{int}} - \mathbf{r}^{\text{src}}) + \sum_{p' =
    1}^{P} G_{0}( \mathbf{r}^{\text{int}}_{p} - ( \mathbf{r}^{\text{int}}_{p'} -
  \delta^{\text{int}} \hat{\mathbf{z}} ) ) a_{p'}\\
  = \sum_{p' = 1}^{P} G_{1}( \mathbf{r}_{p}^{\text{int}} -
  ( \mathbf{r}^{\text{int}}_{p'} + \delta^{\text{int}} \hat{\mathbf{z}} ) ) b_{p'} \\
  + \sum_{q = 1}^{Q} G_{1}( \mathbf{r}^{\text{int}}_{p} - ( \mathbf{r}^{\text{tar}}_{q} -
  \delta^{\text{tar}} \hat{\boldsymbol{\nu}}_{q} ) ) c^{\text{ext}}_{q},
  \label{eq:Interface1}
\end{multline}
and
\begin{multline}
  \partial_{n} G_{0}(\mathbf{r}_{p}^{\text{int}} - \mathbf{r}^{\text{src}}) +
  \sum_{p' = 1}^{P} \partial_{n} G_{0}( \mathbf{r}^{\text{int}}_{p} - (
  \mathbf{r}^{\text{int}}_{p'} - \delta^{\text{int}} \hat{\mathbf{z}} ) ) a_{p'}\\
  = \epsilon_{r1}^{-1} \sum_{p' = 1}^{P} \partial_{n} G_{1}(
  \mathbf{r}_{p}^{\text{int}} -
  ( \mathbf{r}^{\text{int}}_{p'} + \delta^{\text{int}} \hat{\mathbf{z}} ) ) b_{p'}
  \\
  + \epsilon_{r1}^{-1} \sum_{q = 1}^{Q} \partial_{n} G_{1}(
  \mathbf{r}^{\text{int}}_{p} - ( \mathbf{r}^{\text{tar}}_{q} - \delta^{\text{tar}}
  \hat{\boldsymbol{\nu}}_{q} ) ) c^{\text{ext}}_{q},
  \label{eq:Interface2}
\end{multline}
for $p = 1, \dots, P$. Note that
$\partial_{n} = \hat{\mathbf{n}}_{p} \cdot \nabla$ in the notation
above. Equations \eqref{eq:Interface1} and \eqref{eq:Interface2} give
$2P$ conditions.

For boundary conditions \eqref{eq:TargetBCs}, we require
\begin{multline}
  \sum_{p = 1}^{P} G_{1}( \mathbf{r}^{\text{tar}}_{q} -
  ( \mathbf{r}^{\text{int}}_{p} + \delta^{\text{int}} \hat{\mathbf{z}}
  ) ) b_{p} \\
  + \sum_{q' = 1}^{Q} G_{1}( \mathbf{r}^{\text{tar}}_{q} - (
  \mathbf{r}^{\text{tar}}_{q'} -  \delta^{\text{tar}}
  \hat{\boldsymbol{\nu}}_{q'} ) ) c^{\text{ext}}_{q'}\\
  =
  \sum_{q' = 1}^{Q} G_{2}( \mathbf{r}^{\text{tar}}_{q} -
  ( \mathbf{r}^{\text{tar}}_{q'} + \delta^{\text{tar}}
  \hat{\boldsymbol{\nu}}_{q'} ) ) c^{\text{int}}_{q'},
  \label{eq:TargetBdy1}
\end{multline}
and
\begin{multline}
  \epsilon_{r1}^{-1} \sum_{p = 1}^{P} \partial_{\nu} G_{1}(
  \mathbf{r}^{\text{tar}}_{q} -
  ( \mathbf{r}^{\text{int}}_{p} + \delta^{\text{int}} \hat{\mathbf{z}}
  ) ) b_{p} \\
  + \epsilon_{r1}^{-1} \sum_{q' = 1}^{Q} \partial_{\nu} G_{1}(
  \mathbf{r}^{\text{tar}}_{q} - ( \mathbf{r}^{\text{tar}}_{q'} -
  \delta^{\text{tar}} \hat{\boldsymbol{\nu}}_{q'} ) )
  c^{\text{ext}}_{q'}\\
  = \epsilon_{r2}^{-1} \sum_{q' = 1}^{Q} \partial_{\nu} G_{2}(
  \mathbf{r}^{\text{tar}}_{q} - ( \mathbf{r}^{\text{tar}}_{q'} +
  \delta^{\text{tar}} \hat{\boldsymbol{\nu}}_{q'} ) )
  c^{\text{int}}_{q'},  
  \label{eq:TargetBdy2}
\end{multline}
for $q = 1, \dots, Q$. Note that
$\partial_{\nu} = \hat{\boldsymbol{\nu}}_{q} \cdot \nabla$ in the
notation above. Equations \eqref{eq:TargetBdy1} and
\eqref{eq:TargetBdy2} give $2Q$ equations.

\subsection{MFS system}

Let $A_{1}$, $A_{2}$, $A_{3}$ and $A_{4}$ be $P \times P$ matrices
whose entries are given by
\begin{subequations}
  \begin{align}
    [ A_{1} ]_{pp'} &= G_{0}( \mathbf{r}^{\text{int}}_{p} - (
    \mathbf{r}^{\text{int}}_{p'} - \delta^{\text{int}}
                      \hat{\mathbf{z}} ) ),\\
    [ A_{2} ]_{pp'} &= G_{1}( \mathbf{r}^{\text{int}}_{p} - (
    \mathbf{r}^{\text{int}}_{p'} + \delta^{\text{int}}
                      \hat{\mathbf{z}} ) ),\\
    [ A_{3} ]_{pp'} &= \partial_{n} G_{0}( \mathbf{r}^{\text{int}}_{p} - (
    \mathbf{r}^{\text{int}}_{p'} - \delta^{\text{int}}
                      \hat{\mathbf{z}} ) ),\\
    [ A_{4} ]_{pp'} &= \epsilon_{r1}^{-1} \partial_{n} G_{1}(
                      \mathbf{r}^{\text{int}}_{p} - (
    \mathbf{r}^{\text{int}}_{p'} + \delta^{\text{int}} \hat{\mathbf{z}} ) ).
  \end{align}
  \label{eq:A-blocks}
\end{subequations}
Let $B_{1}$ and $B_{2}$ be the $P \times Q$ matrices whose entries are
given by
\begin{subequations}
  \begin{align}
    [ B_{1} ]_{pq} &= G_{1}( \mathbf{r}^{\text{int}}_{p} - (
    \mathbf{r}^{\text{tar}}_{q} - \delta^{\text{tar}}
                     \hat{\boldsymbol{\nu}}_{q} ) ),\\
    [ B_{2} ]_{pq} &= \epsilon_{r1}^{-1} \partial_{n} G_{1}(
                     \mathbf{r}^{\text{int}}_{p} - (
    \mathbf{r}^{\text{tar}}_{q} - \delta^{\text{tar}}
                     \hat{\boldsymbol{\nu}}_{q} ) ).
  \end{align}
\end{subequations}
Let $C_{1}$ and $C_{2}$ be the $Q \times P$ matrices whose entries are
given by
\begin{subequations}
  \begin{align}
    [ C_{1} ]_{qp} &= G_{1}( \mathbf{r}^{\text{tar}}_{q} - (
    \mathbf{r}^{\text{int}}_{p} + \delta^{\text{int}}
                     \hat{\mathbf{n}}_{p} ) ),\\
    [ C_{2} ]_{qp} &= \epsilon_{r1}^{-1} \partial_{\nu} G_{1}(
                     \mathbf{r}^{\text{tar}}_{q} - (
    \mathbf{r}^{\text{int}}_{p} + \delta^{\text{int}}
                     \hat{\mathbf{n}}_{p} ) ).
  \end{align}
\end{subequations}
Finally, let $S_{1}$, $S_{2}$, $S_{3}$, and $S_{4}$ denote the $Q
\times Q$ matrices whose entries are given by
\begin{subequations}
  \begin{align}
    [ S_{1} ]_{qq'} &= G_{1}( \mathbf{r}^{\text{tar}}_{q} - (
    \mathbf{r}^{\text{tar}}_{q'} - \delta^{\text{tar}}
                      \hat{\boldsymbol{\nu}}_{q'} ) ),\\
    [ S_{2} ]_{qq'} &= G_{2}( \mathbf{r}^{\text{int}}_{q} - (
    \mathbf{r}^{\text{tar}}_{q'} + \delta^{\text{tar}}
                      \hat{\boldsymbol{\nu}}_{q'} ) ),\\
    [ S_{3} ]_{qq'} &= \epsilon_{r1}^{-1} \partial_{n} G_{1}(
                      \mathbf{r}^{\text{tar}}_{q} - (
    \mathbf{r}^{\text{tar}}_{q'} - \delta^{\text{tar}}
                      \hat{\boldsymbol{\nu}}_{q'} ) ),\\
    [ S_{4} ]_{qq'} &= \epsilon_{r2}^{-1} \partial_{n} G_{2}(
                      \mathbf{r}^{\text{tar}}_{q} - (
    \mathbf{r}^{\text{tar}}_{q'} + \delta^{\text{tar}}
                      \hat{\boldsymbol{\nu}}_{q'} ) ).
  \end{align}
\end{subequations}
Because $\delta^{\text{int}} \neq 0$ and $\delta^{\text{tar}} \neq 0$,
these matrix entries never correspond to evaluation of the free-space
Green's function where it is singular.

Using these matrices, we find that the $2P + 2Q$ equations given by
\eqref{eq:Interface1}, \eqref{eq:Interface2}, \eqref{eq:TargetBdy1},
and \eqref{eq:TargetBdy2} as the following block linear system,
\begin{equation}
  \begin{bmatrix}
    -A_{1} & A_{2} & B_{1} & \\
    -A_{3} & A_{4} & B_{2} & \\
    & C_{1} & S_{1} & -S_{2} \\
    & C_{2} & S_{3} & -S_{4}
  \end{bmatrix}
  \begin{bmatrix}
    \mathbf{a} \\ \mathbf{b} \\ \mathbf{c}^{\text{ext}} \\
    \mathbf{c}^{\text{int}}
  \end{bmatrix}
  =
  \begin{bmatrix}
    \mathbf{y}_{1} \\ \mathbf{y}_{2} \\ 0 \\ 0
  \end{bmatrix}.
  \label{eq:block-system}
\end{equation}
Here, $\mathbf{a}$ and $\mathbf{b}$ are the $P$-vectors whose
components are $a_{p}$ and $b_{p}$ for $p = 1, \dots, P$,
respectively, and $\mathbf{c}^{\text{ext}}$ and
$\mathbf{c}^{\text{int}}$ are the $Q$-vectors whose components are
$c^{\text{ext}}_{q}$ and $c^{\text{int}}_{q}$ for $q = 1, \dots, Q$,
respectively. The right-hand side is made up of $P$-vectors
$\mathbf{y}_{1}$ and $\mathbf{y}_{2}$ whose components are given by
\begin{subequations}
  \begin{align}
    [ \mathbf{y}_{1} ]_{p} &= G_{0}( \mathbf{r}^{\text{int}}_{p} -
                             \mathbf{r}^{\text{src}} ),\\
    [ \mathbf{y}_{2} ]_{p} &= \partial_{n} G_{0}(
                             \mathbf{r}^{\text{int}}_{p} -
                             \mathbf{r}^{\text{src}} ),
  \end{align}
\end{subequations}
for $p = 1, \dots, P$.

\subsection{Example}

Block system \eqref{eq:block-system} can be solved numerically leading
to the determination of the coefficients in
\eqref{eq:MFS0}, \eqref{eq:MFS1}, and \eqref{eq:MFS2}. With those
expansion coefficients determined, those approximations can be
evaluated anywhere within the regions to which they are defined. Here,
we show an example computation.

\begin{figure}[htb]
  \centering
  \includegraphics[width=0.8\linewidth]{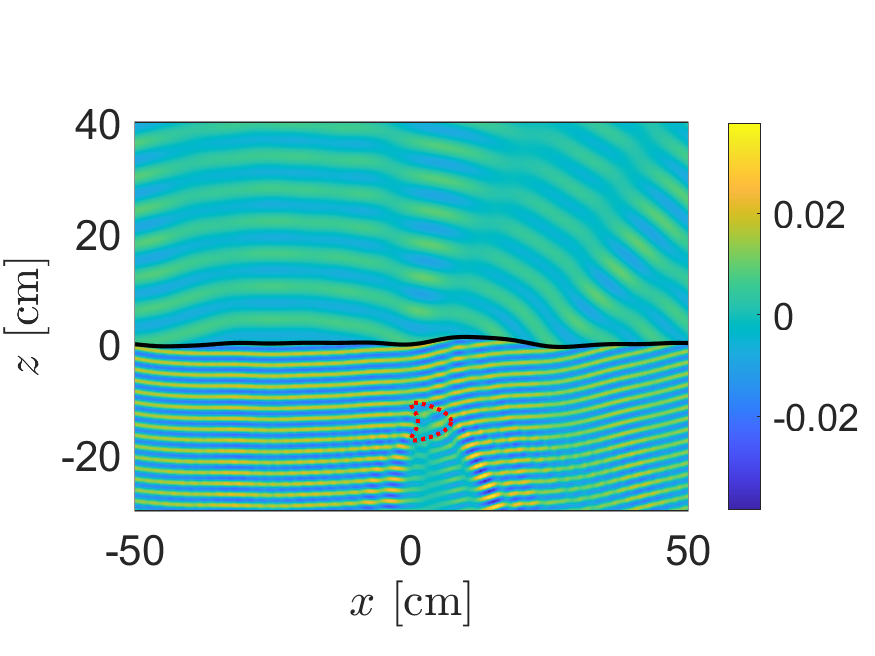}\\
  \vspace{-6pt}
  \includegraphics[width=0.8\linewidth]{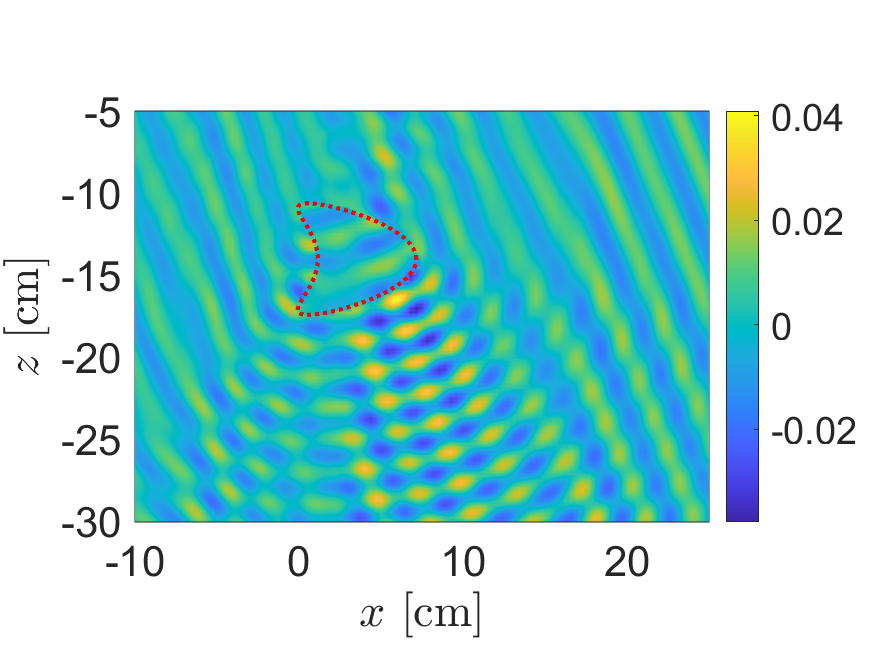}
  \caption{Example MFS result for the direct scattering problem. The
    upper plot shows the scattered fields over a larger region that
    includes the rough air-soil interface and the target. The lower
    plot shows the scattered fields in a smaller region about the
    target. The air-soil interface is plotted as a solid black curve
    and the kite-shaped target is plotted as a dotted red curve.}
  \label{fig:forward-example}
\end{figure}

In Fig.~\ref{fig:forward-example} we show MFS results for the fields
scattered due to a point source located at
$\mathbf{r}_{\text{src}} = (-25,75)$ cm, at frequency $f = 5.1$ GHz
and with $c = 30$ cm GHz. For soil, we have used $\epsilon_{r1} = 9$
and for the target we have used $\epsilon_{r2} = 5$.

The air-soil interface is one realization of a Gaussian-correlated
random rough surface with RMS height $h_{\text{RMS}} = 0.4$ cm and
correlation length $\ell = 8$ cm generated using the method described
by Tsang et al~\cite{tsang2004scattering}. In
Fig.~\ref{fig:forward-example}, we have used $L = 400$ cm and
$P = 512$ for the air-soil interface. For the target, we consider the
``kite'' whose boundary is given by
$\mathbf{r}^{\text{tar}}(t) = ( 3, -14 ) + ( 3.0 \cos t + 1.8 \cos 2t
- 0.65, 3.4 \sin t )$ and $Q = 128$. We have set
$\delta^{\text{int}} = \delta^{\text{tar}} = 0.1$.

The example shown in Fig.~\ref{fig:forward-example} illustrates that
the MFS approximation captures the complex interactions of the fields
with the rough air-soil interface and the target. These complex
interactions involve a variety of spatial scales and the computed
solution smoothly transitions over these scales. Because the MFS is an
easy to implement method and accurately accounts for the underlying
physics of multiple scattering by the rough air-soil interface and the
target, we use it for simulating measurements.

\section{The inverse scattering problem}
\label{sec:inverse-problem}

The objective for the inverse scattering problem is to reconstruct the
subsurface target from GP-SAR measurements. That target is
characterized completely by its ``support'' given by $\Omega$ and its
relative dielectric constant $\epsilon_{r2}$. The extent to which we
can recover all or some of these characteristics depend on the
information contained in measurements. For GP-SAR measurements, we
identify inherent limitations that affect the solution of the inverse
scattering problem.

\subsection{GP-SAR measurements}

GP-SAR measurements are multi-frequency signals taken over several
locations along the flight path. Let $\omega_{m}$ for
$m = 1, \dots, M$ denote the $M$ frequencies used to sample the system
bandwidth and let $\mathbf{r}_{n}$ for $n = 1, \dots, N$ denote the
$N$ locations along the flight path where these multi-frequency
signals are emitted and measured. For each location, we set
$\mathbf{r}_{\text{src}} = \mathbf{r}_{n}$ and solve the direct
scattering problem described above for frequency $\omega_{m}$. We take
as measurements evaluations of $u_{0}(\mathbf{r}_{n};\omega_{m})$
corresponding to the scattered field $u_{0}$ evaluated at frequency
$\omega_{m}$ at position $\mathbf{r}_{n}$. Through repeated solution
of the direct scattering problem over all $M$ frequencies and all $N$
source/measurement positions, we obtain the measurement matrix
$D \in \mathbb{C}^{M \times N}$.  Measurements are typically corrupted
by additive noise, so we model the entries of the data matrix as the
sum,
\begin{equation}
  d_{mn} = u_{0}(\mathbf{r}_{n};\omega_{m}) + \eta_{mn}
  \label{eq:SAR-measurements}
\end{equation}
with $\eta_{mn}$ denoting additive Gaussian white noise.

Note that we have considered only frequency-flat, isotropic point
sources. Additionally, we use point-wise evaluations of the scattered
field as measurements.  Different sources and measurements can be used
through an appropriate modification of the direct scattering
problem. However, the measurement matrix we use here is fundamental in
that other sources and measurements can be computed from it.

\subsection{Principal component analysis for ground bounce removal}

A key challenge in this inverse scattering problem is the
ground bounce signals in the measurements. Ground bounce signals
correspond to the portion of measurements that are reflections
of the incident field by the air-soil interface. They contribute more
to the overall power in measurements than signals scattered by the
subsurface target. However, those signals carry little or no
information about the target. Therefore, we must remove ground bounce
signals from measurements to be able to image the subsurface target.

When the air-soil interface is known, one can model these reflections
and use those to remove them from measurements. Otherwise, one needs
to consider alternative methods. In what follows, we describe how to
use principal component analysis (PCA) to approximately remove ground
bounce signals from measurements.

For PCA, we first compute the singular value decomposition (SVD) of
the data matrix so that $D = U \Sigma V^{H}$ where $[\cdot]^{H}$
denotes the Hermitian (conjugate) transpose. We assume the non-zero
singular values $\sigma_{1}, \sigma_{2}, \dots, \sigma_{r}$ along the
diagonal of $\Sigma$ are in descending order. The key idea with PCA is
that ground bounce signals are stronger than the scattered signals, so
they are associated with the largest singular values. Hence, we
consider
\begin{equation}
  \tilde{D}_{J} = D - \sum_{j = 1}^{J} \sigma_{j} \mathbf{u}_{j}
  \mathbf{v}_{j}^{H}
  \label{eq:PCA}
\end{equation}
instead of $D$. Here, $J$ is a user-defined truncation,
$\mathbf{u}_{j}$ and $\mathbf{v}_{j}$ are the $j$th columns of $U$ and
$V$, respectively. Provided that $J$ is chosen well, the resulting
processed data matrix $\tilde{D}_{J}$ is amenable to imaging methods such
as Kirchhoff migration which we explain below.

Using PCA to approximately remove ground bounce signals is completely
data-driven. It does not require {\it a priori} knowledge of the
electromagnetic properties of soil. Rather, it is based off of the
assumption that the ground bounce signals dominate the measurements
over the scattered signals by the target.

\subsection{Kirchhoff migration}

Kirchhoff migration (KM), also known as back-propagation or reverse
time-migration, is a so-called sampling
method~\cite{potthast2006survey}. Rather than having to solve an
optimization problem to reconstruct an image of the target, KM
requires only the evaluation of an imaging function over some region
of interest, which we call the imaging region. Let $\mathbf{y}$ denote
a point in this imaging region. The KM imaging function is given by
\begin{equation}
  I^{\text{KM}}(\mathbf{y}) = \sum_{m = 1}^{M} \sum_{n = 1}^{N}
  \tilde{d}_{mn} a_{mn}^{\ast}(\mathbf{y}),
  \label{eq:KM-function}
\end{equation}
with $\tilde{d}_{mn}$ denoting the entry of $\tilde{D}_{J}$,
$a_{mn}(\mathbf{y})$ denoting what we call the illuminations and
$[\cdot]^{\ast}$ denoting the complex conjugate.

To define the illuminations, we assume we know the mean elevation $L$
of the flight path over the air-soil interface. Additionally, we
assume we know the relative refractive index for soil denoted by
$\epsilon_{r1}$. Let $\mathbf{r}_{n} = ( x_{n}, L )$,
$\mathbf{y} = ( \xi, \zeta )$, and $k_{m} = \omega_{m}/c$.  The
illuminations are given by
\begin{equation}
  a_{mn}(\mathbf{y}) = e^{\mathrm{i} 2 k_{m} L (1 + ( x_{n} - \xi
    )^{2}/2L^{2})} e^{-\mathrm{i} 2 k_{m} \sqrt{\epsilon_{r1}} \zeta}.
  \label{eq:illuminations}
\end{equation}
The illuminations given in \eqref{eq:illuminations} correspond to the
phase accumulated from scattering by a point target at location
$\mathbf{y}$ below a {\em flat} air-soil interface on $z = 0$ using
the Fresnel approximation. Additionally, illuminations
\eqref{eq:illuminations} assume only one interaction between this
point target and the flat air-soil interface, whereas the full
boundary value problem includes infinitely-many of these interactions.

We form an image by plotting $|I^{\text{KM}}|$ over the imaging
region. When KM is effective, this plot reveals spatial information
about the target. Note that this implementation of KM starts with
measurements defined by \eqref{eq:SAR-measurements} which include
additive measurement noise. We then compute \eqref{eq:PCA}, the
processed data matrix $\tilde{D}$ resulting from PCA which
approximately removes ground bounce signals. Finally, this KM
implementation evaluates \eqref{eq:KM-function} with illuminations
\eqref{eq:illuminations} defined for a flat air-soil interface.

In this formulation of KM, we assume {\it a priori} knowledge of a
ground permittivity $\epsilon_{r1}$ and the mean height of the flight
path over the air-soil interface $L$. Other than those two parameters,
no other parameters are needed to evaluate \eqref{eq:KM-function}.

\subsection{Examples}

We now show examples of images produced using the methods described
above on data simulated using the MFS. For these examples, we have
used $41$ frequencies uniformly sampling the bandwidth between $3.5$
GHz and $5.5$ GHz (at $50$ MHz steps) similarly to what was done by
Garcia-Fernandez {\it et al}~\cite{garcia2019bistatic}. We have set
the elevation above the mean of the air-soil interface to be $L = 75$
cm and use $35$ spatial locations unformly sampling the synthetic
aperture of size $a = 102$ cm (at $3$ cm steps).

We consider the air-soil interface to be one realization of a
Gaussian-correlated random rough surface with RMS height $0.4$ cm and
correlation length $8$ cm. We showed recently that random rough
surfaces with these parameters exhibit enhanced backscattering
indicating significant multiple scattering caused by the surface
roughness.  As suggested by Daniels~\cite{daniels:2006}, we set
$\epsilon_{r1} = 9$ for soil and $\epsilon_{r2} = 2.3$ for the
target. For the MFS computations to generate simulated measurements,
we used $P = 512$ points to sample the surface over the interval of
length $L = 400$ cm.

\begin{figure}[htb]
  \centering
  \begin{subfigure}[t]{0.48\linewidth}
    \includegraphics[width=\linewidth]{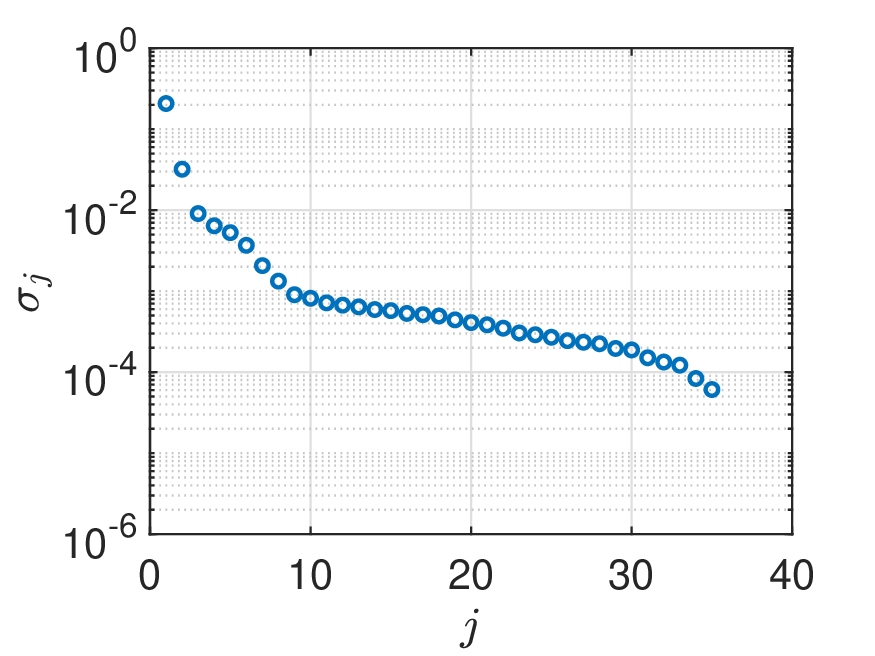}
    \caption{Singular values of $D$}
  \end{subfigure}
  \begin{subfigure}[t]{0.48\linewidth}
    \includegraphics[width=\linewidth]{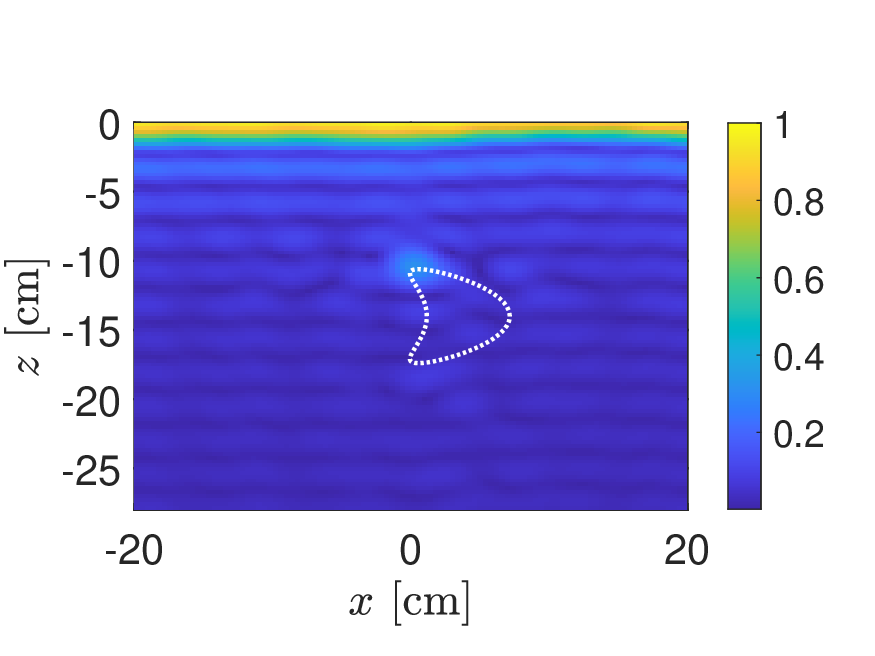}
    \caption{KM image using $D$}
  \end{subfigure}
  \begin{subfigure}[t]{0.48\linewidth}
    \includegraphics[width=\linewidth]{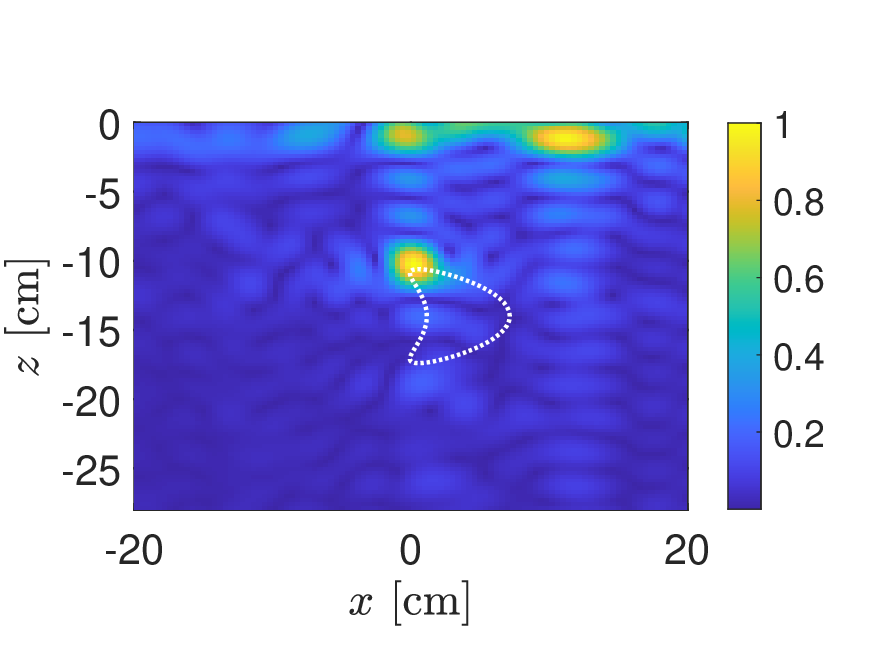}
    \caption{KM image using $\tilde{D}_{1}$}
  \end{subfigure}
  \begin{subfigure}[t]{0.48\linewidth}
    \includegraphics[width=\linewidth]{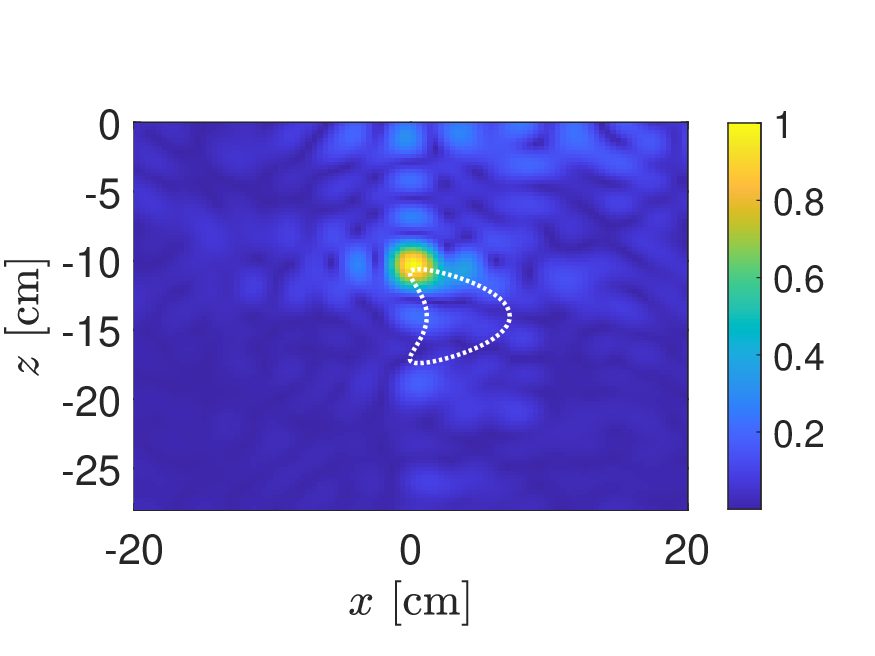}
    \caption{KM image using $\tilde{D}_{2}$}
  \end{subfigure}
  \caption{(a) Singular values of the data matrix $D$ with measurement
    noise added so that SNR $\approx 25$ dB. (b) KM image produced
    using KM on $D$. (c) KM image produced using KM on
    $\tilde{D}_{1}$. (d) KM image produced using KM on
    $\tilde{D}_{2}$. All KM images show $| I^{\text{KM}}|$ plotted
    over the imaging region normalized by their respective maximum
    values. }
  \label{fig:PCA+KM}  
\end{figure}

In Fig.~\ref{fig:PCA+KM} we show results for the same ``kite'' target
used in Fig.~\ref{fig:forward-example}. We have added measurement noise
to the simulated measurements so that SNR $\approx 25$ dB. The
singular values of the data matrix $D$ are shown in
Fig.~\ref{fig:PCA+KM}(a). We observe from those results that the first
two singular values are significantly larger than the remaining
singular values with large gaps between them. In
Figs.~\ref{fig:PCA+KM}(b)--(d) we show KM images applied to the
original data matrix $D$ (Fig.~\ref{fig:PCA+KM}(b)), the PCA processed
data matrix $\tilde{D}_{1}$ with one term removed
(Fig.~\ref{fig:PCA+KM}(c)), and the PCA processed data matrix
$\tilde{D}_{2}$ with two terms removed (Fig.~\ref{fig:PCA+KM}(d)). These
KM images show plots of the absolute value of \eqref{eq:KM-function}
normalized by its maximum value.

These results show how PCA effectively removes the ground bounce
signals that interfere with the identification and location of the
target. In Fig.~\ref{fig:PCA+KM}(b) we find that the image is
concentrated near $z = 0$ corresponding to the air-soil
interface. With the contribution by the first singular value removed,
the image shown in Fig.~\ref{fig:PCA+KM}(c) identifies the target but
also has artifacts located near the air-soil interface. After removing
the contributions by the first two singular values, we find in
Fig.~\ref{fig:PCA+KM}(d) that the resulting image identifies the
target. Specifically, it appears to identify the single point on the
boundary of the target that is closest to the synthetic aperture.

\begin{figure}[htb]
  \centering
  \begin{subfigure}{0.48\linewidth}
    \includegraphics[width=\linewidth]{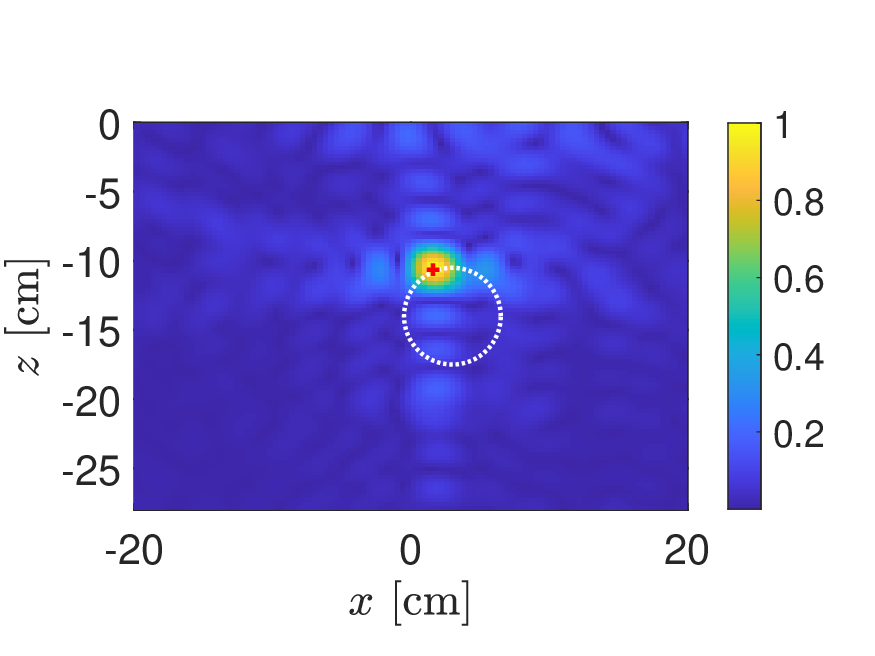}
    \caption{circular disk}
  \end{subfigure}
  \begin{subfigure}{0.48\linewidth}
    \includegraphics[width=\linewidth]{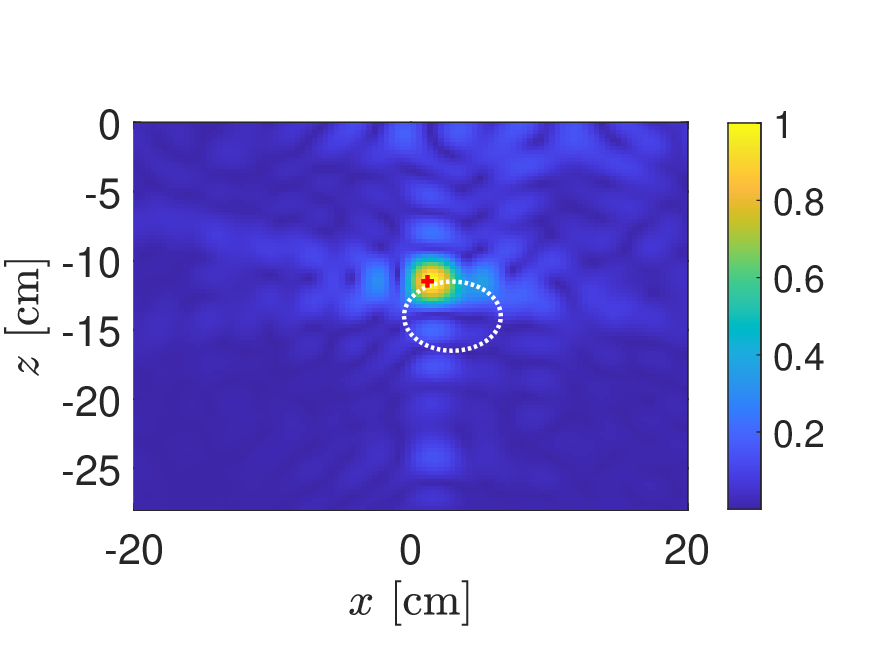}
    \caption{ellipse}
  \end{subfigure}
  \begin{subfigure}{0.48\linewidth}
    \includegraphics[width=\linewidth]{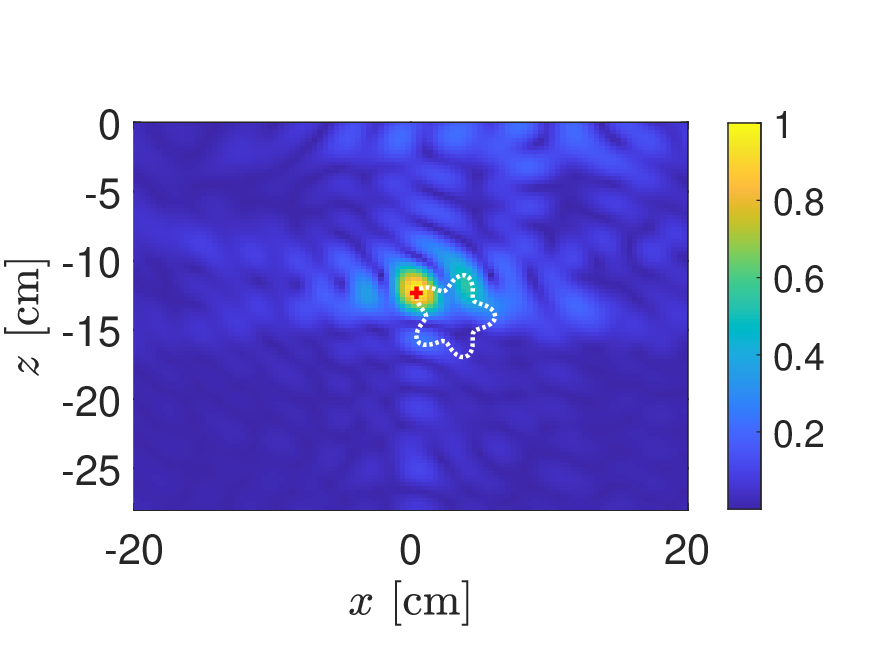}
    \caption{star}
  \end{subfigure}
  \begin{subfigure}{0.48\linewidth}
    \includegraphics[width=\linewidth]{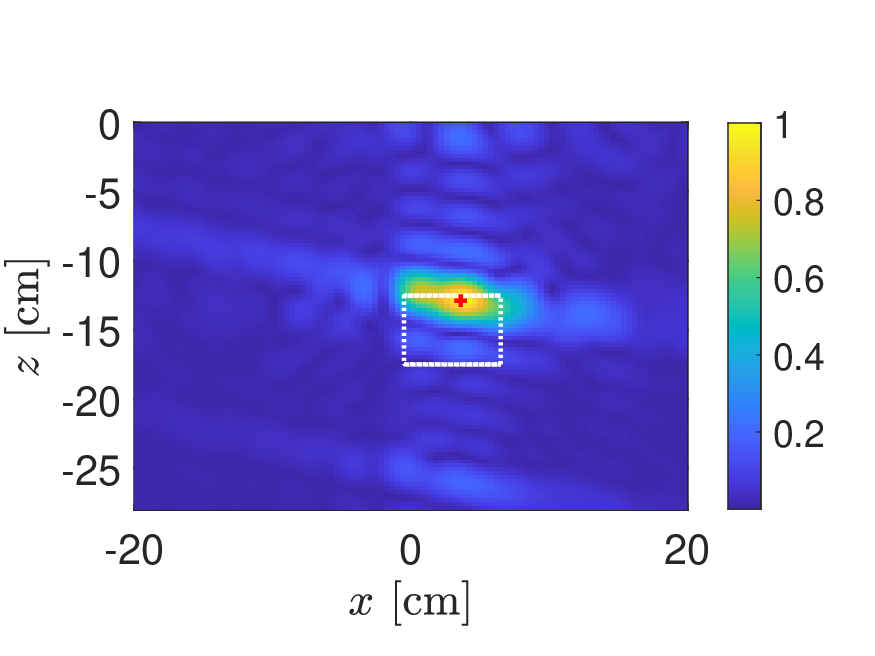}
    \caption{rectangle}
  \end{subfigure}
  \caption{KM images for different shaped targets. For all images, we
    have used $\tilde{D}_{2}$ and SNR $\approx 25$ dB. All KM images
    show $| I^{\text{KM}}|$ plotted over the imaging region normalized
    by their respective maximum values. The location where
    $|I^{\text{KM}}|$ attains its maximum value is plotted as a red
    ``+'' symbol.}
  \label{fig:KMshapes}  
\end{figure}

In Fig.~\ref{fig:KMshapes} we consider the KM images for differently
shaped targets: (a) a circle of radius $3.5$ cm centered at $(3,-14)$
cm, (b) an ellipse with major axis $3.5$ cm and minor axis $2.5$ cm
centered at $(3,-14)$ cm, (c) a star whose boundary is given by
$(3,-14) + (2.5 + 0.6 \cos 5 t, -3 \sin 5 t)$ cm for
$0 \le t \le 2 \pi$, and (d) a rectangle with width $7$ cm and height
$5$ cm centered at $(3,-15)$ cm. All of the targets have relative
dielectric constant $\epsilon_{r2} = 2.3$. All of these images have
used PCA processed data $\tilde{D}_{2}$ and these KM images show plots
of the absolute value of \eqref{eq:KM-function} normalized by its
maximum value.

The maximum value of each of the KM images in
Fig.~\ref{fig:KMshapes}(a)--(d) are plotted as a red ``+'' symbol.
All of these results indicate that the KM image concentrates on a
single point near the boundary of the target facing the synthetic
aperture. There are some differences of the spread of the KM image
about that point most notably for the rectangle. Nonetheless, these KM
images do no provide too much spatial information other than this
point. This behavior of KM images is due to the limitations in
measurements. Specifically, SAR measurements
\eqref{eq:SAR-measurements} only subtend a relatively small portion of
all scattering angles. Moreover, monostatic SAR measurements are
restricted to just the backscattering or retroreflected
direction. These space/angle limitations correspond to a severely
limited aperture imaging problem. We cannot expect to recover very
much spatial information about the target. The images shown in
Fig.~\ref{fig:KMshapes} suggest that one cannot obtain much more than
a single point associated with a target. This point can be used to
identify a target and locate it, but not much else.

\section{Measurement model}
\label{sec:model}

The results shown in the examples above suggest that there may be a
simplified model for measurements. In what follows, we identify, test,
and validate simplifying assumptions leading to a simplified model. We
then propose how to use this simplified model for quantitative imaging
of subsurface targets.

\subsection{First-order target-interface interactions}

The first simplifying assumption we make is to consider only
first-order interactions between the target and the air-soil
interface. Even when there is multiple scattering in the direct
scattering problem, methods for the inverse scattering problem based
only on first-order interactions are effective. It is understood that
these first-order interaction approximations carry sufficient phase
information regarding boundaries where discontinuities in wave fields
occur.

Assuming only first-order interactions between the target and the
air-soil interface, we can consider the following procedure to
approximate the solution of the direct scattering problem.
\begin{enumerate}

\item For a source located above the air-soil interface, compute
  reflection and transmission by the rough air-soil interface with no
  target below it.

\item Use the field transmitted across the air-soil interface to
  determine the field incident on the target.

\item Solve the scattering problem by the target using that incident
  field.

\item Compute the field scattered by the target that is transmitted
  back across the air-soil interface from soil into air and evaluate
  that result at the receiver.

\end{enumerate}
Using block system \eqref{eq:block-system} for the MFS implementation
we have used here, this procedure corresponds to first solving
\begin{equation}
  \begin{bmatrix}
    -A_{1} & A_{2} \\
    -A_{3} & A_{4}  \\
  \end{bmatrix}
  \begin{bmatrix}
    \mathbf{a}^{(0)} \\ \mathbf{b}^{(0)}
  \end{bmatrix}
  =
  \begin{bmatrix}
    \mathbf{y}_{1} \\ \mathbf{y}_{2}
  \end{bmatrix},
  \label{eq:approx-step1}
\end{equation}
then solving
\begin{equation}
  \begin{bmatrix}
    S_{1} & -S_{2} \\
    S_{3} & -S_{4}
  \end{bmatrix}
  \begin{bmatrix}
    \mathbf{c}^{\text{ext}} \\ \mathbf{c}^{\text{int}}
  \end{bmatrix}
  = -
  \begin{bmatrix}
    C_{1} \mathbf{b}^{(0)} \\ C_{2} \mathbf{b}^{(0)}
  \end{bmatrix},
  \label{eq:approx-step2}
\end{equation}
and finally solving
\begin{equation}
  \begin{bmatrix}
    -A_{1} & A_{2} \\
    -A_{3} & A_{4}  \\
  \end{bmatrix}
  \begin{bmatrix}
    \mathbf{a}^{(1)} \\ \mathbf{b}^{(1)}
  \end{bmatrix}
  = -
  \begin{bmatrix}
    B_{1} \mathbf{c}^{\text{ext}} \\ B_{2} \mathbf{c}^{\text{ext}}
  \end{bmatrix}.
  \label{eq:approx-step3}
\end{equation}
With these results, we approximate measurements through evaluation of
\eqref{eq:MFS0} using $a_{p} \approx a_{p}^{(0)} + a_{p}^{(1)}$.

\begin{figure}[htb]
  \centering
  \includegraphics[width=0.85\linewidth]{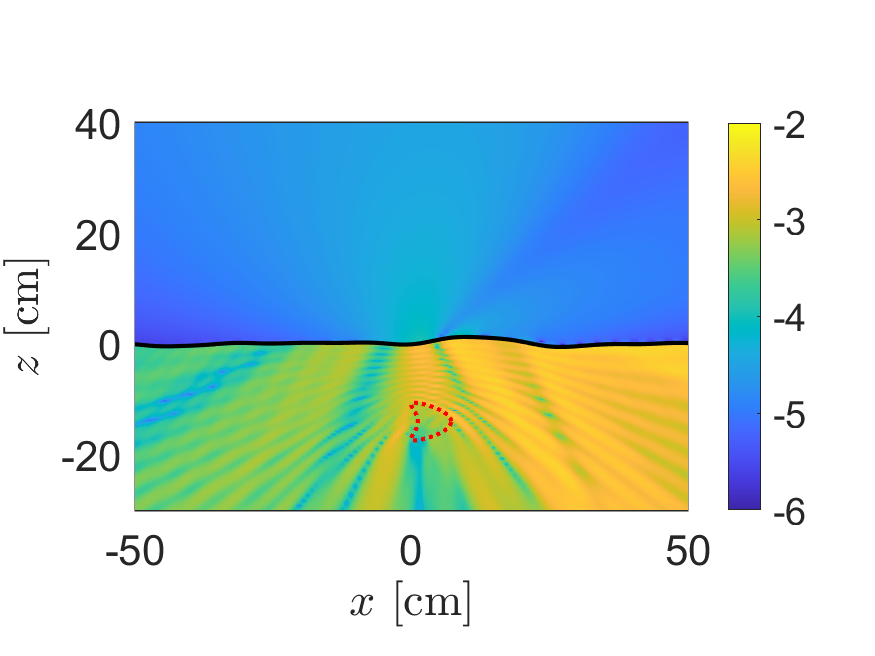}
  \caption{Absolute difference ($\log_{10}$-scale) between the
    solution of the full boundary value problem and the first-order
    target-interface interaction approximation. The kite-shaped target
    used for both computations is plotted as a dotted red curve.}
  \label{fig:first-order}
\end{figure}

In Fig.~\ref{fig:first-order} we show the absolute difference
($\log_{10}$-scale) between the MFS solution of the full boundary
value problem described in Section \ref{sec:MFS} and the first-order
target-interface interaction approximation described here. For these
results, we have used one realization of a Gaussian-correlated random
rough surface with RMS height $0.4$ cm and correlation length $8$
cm. The relative dielectric constant for soil is $\epsilon_{r1} = 9$
and for the target is $\epsilon_{r2} = 2.3$. We used a point source at
location $\mathbf{r}_{\text{src}} = ( -25, 75 )$ cm at frequency $5.1$
GHz.

This result is characteristic of this first-order target-interface
interaction approximation. The absolute difference is nearly three
orders of magnitude smaller above the air-soil interface than below
it. This non-uniformity is to be expected since the multiple
interactions between the target and interface that are neglected in
this approximation are most pronounced below the air-soil
interface. However, from the point of view of modeling measurements
taken above the air-soil interface, this result suggest that this
approximation is accurate and useful simplifying the model of SAR
measurements.

\begin{figure}[htb]
  \centering
  \begin{subfigure}{0.85\linewidth}
    \includegraphics[width=\linewidth]{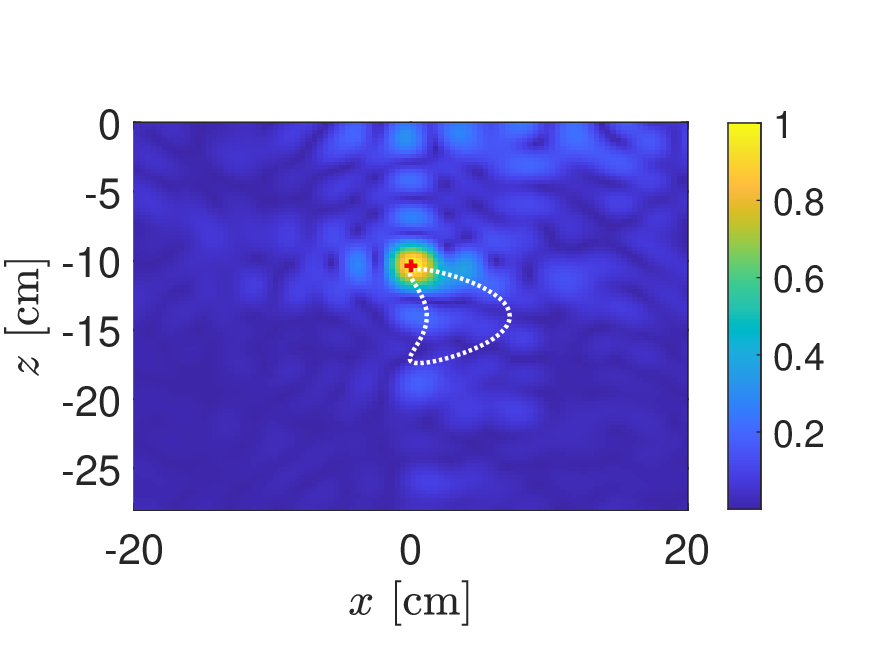}
    \caption{Full problem}
  \end{subfigure}
  \begin{subfigure}{0.85\linewidth}
    \includegraphics[width=\linewidth]{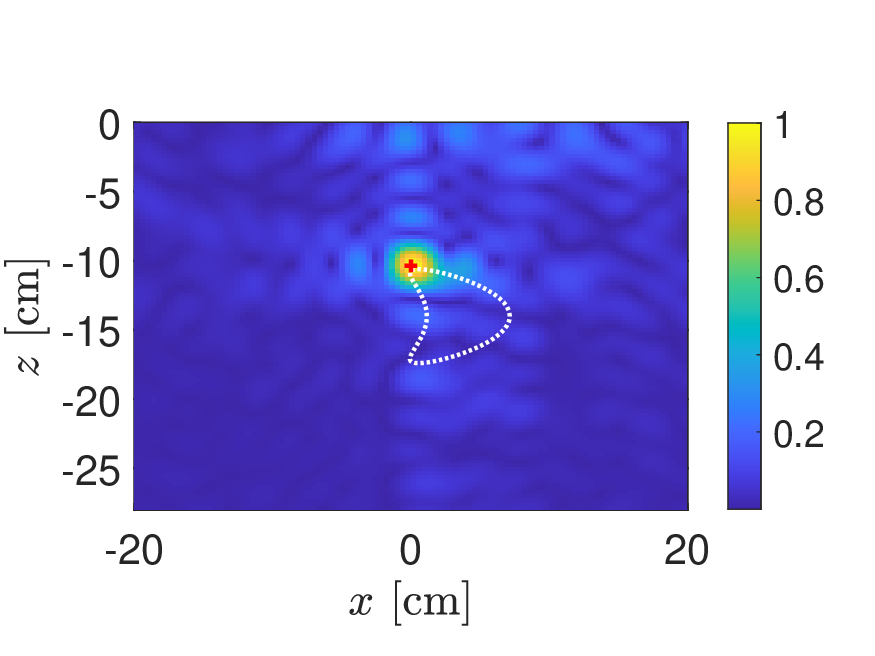}
    \caption{First-order}
  \end{subfigure}
  \caption{Comparison of KM images using simulated measurements of the
    (a) full direct scattering problem and (b) the first-order
    target-interface interaction approximation. For both images we
    have used $\tilde{D}_{2}$ for the respective data matrices and
    both are normalized by their respective maximum values. The
    locations where the KM image attains its peak is plotted as a red
    ``+'' symbol. For both images location of this point is
    $(0,-10.36)$ cm.}
  \label{fig:KM-firstorder}
\end{figure}

In Fig.~\ref{fig:KM-firstorder} we show comparisons of the KM images
produced using measurements computed (a) from the full direct
scattering problem and (b) from the first-order target-interface
interaction approximation. The air-soil interface is the same
realization of a Gaussian-correlated random rough surface with RMS
height $0.4$ cm and correlation length $8$ cm for both data sets.  The
relative dielectric constant in soil is $\epsilon_{r1} = 9$ and in the
the target is $\epsilon_{r2} = 2.3$. Additive measurement noise has
been added so that SNR $\approx 25$ dB.  Both of these images are
produced using $\tilde{D}_{2}$ from their respective data
matrices. These results are nearly indistinguishable from one another
indicating that the first-order approximation is valid for modeling
measurements. The location where the maximum of the KM image is
attained for the first-order approximation is exactly the same as for
for the full direct scattering problem data.

\subsection{Shower curtain effect}

The shower curtain effect is a phenomenon related to imaging through a
scattering medium~\cite{ishimaru, garnier:2024}.  To explain it
consider a source and a detector separated by a fixed distance with
some multiple scattering region of finite extent ({\it i.e.}~the
shower curtain) somewhere in between. The location of the shower
curtain relative to the source affects the image quality. Blurring is
stronger when the shower curtain is closer to the receiver than the
source. When the shower curtain is closer to the source, blurring is
reduced in comparison. Scattering by the shower curtain causes
blurring through introducing angular diversity.  When the receiver is
close to the shower curtain, it measures a substantial fraction of
large-angle scattered fields which, in turn, introduces a mixture of
phases that cause blurring. In contrast, when the receiver is far from
the shower curtain, it does not measure those large-angle scattered
fields thereby reducing blur.

When we use the first-order target-interface approximation described
above, scattering by the target is a secondary source below the rough
air-soil interface which acts as the shower curtain. Since a buried
landmine is typically situated at distances on the order of the
wavelength below the air-soil interface, its distance to it is shorter
than the distance from the UAV to it. Considering the shower curtain
effect, we assume that the roughness of the air-soil interface can be
neglected when computing scattering by the target. For that case, we
solve \eqref{eq:approx-step1} and \eqref{eq:approx-step2} as before,
but replace \eqref{eq:approx-step3} with
\begin{equation}
  \begin{bmatrix}
    -A_{1}^{\text{flat}} & A_{2}^{\text{flat}} \\
    -A_{3}^{\text{flat}} & A_{4}^{\text{flat}}  \\
  \end{bmatrix}
  \begin{bmatrix}
    \tilde{\mathbf{a}}^{(1)} \\ \tilde{\mathbf{b}}^{(1)}
  \end{bmatrix}
  = -
  \begin{bmatrix}
    B_{1} \mathbf{c}^{\text{ext}} \\ B_{2} \mathbf{c}^{\text{ext}}
  \end{bmatrix},
\end{equation}
where the entries of the blocks $A_{1}^{\text{flat}}$,
$A_{2}^{\text{flat}}$, $A_{3}^{\text{flat}}$, and
$A_{4}^{\text{flat}}$ are computed as \eqref{eq:A-blocks} but for a
flat air-soil interface in which $\mathbf{r}_{p}^{\text{int}} = ( x_{p}, 0 )$
for $p = 1, \dots, P$. We call this approximation 
the first-order/flat interface approximation.

\begin{figure}[htb]
  \centering
  \begin{subfigure}{0.85\linewidth}
    \includegraphics[width=\linewidth]{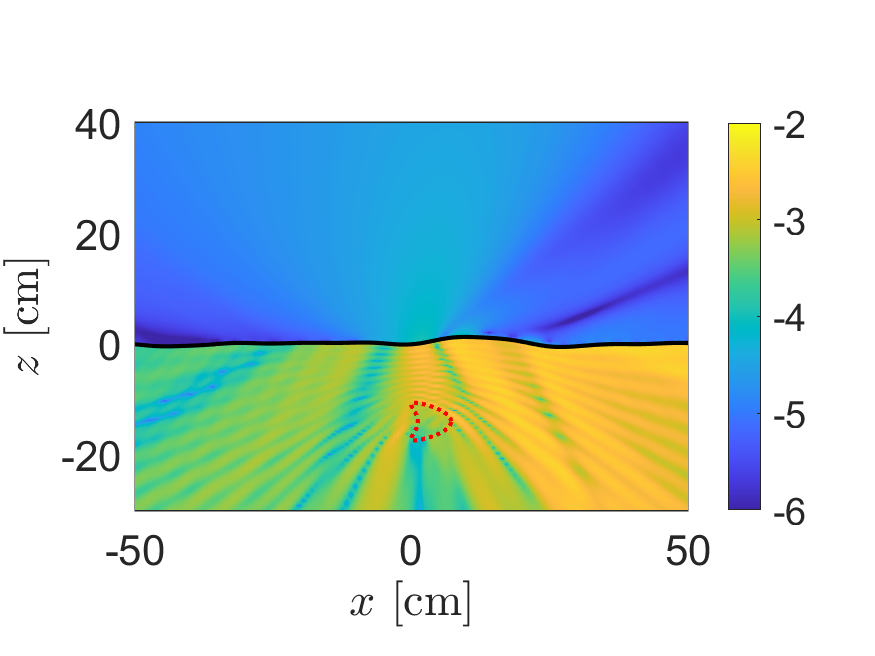}
    \caption{Absolute difference ($\log_{10}$-scale)}
  \end{subfigure}
  \begin{subfigure}{0.85\linewidth}
    \includegraphics[width=\linewidth]{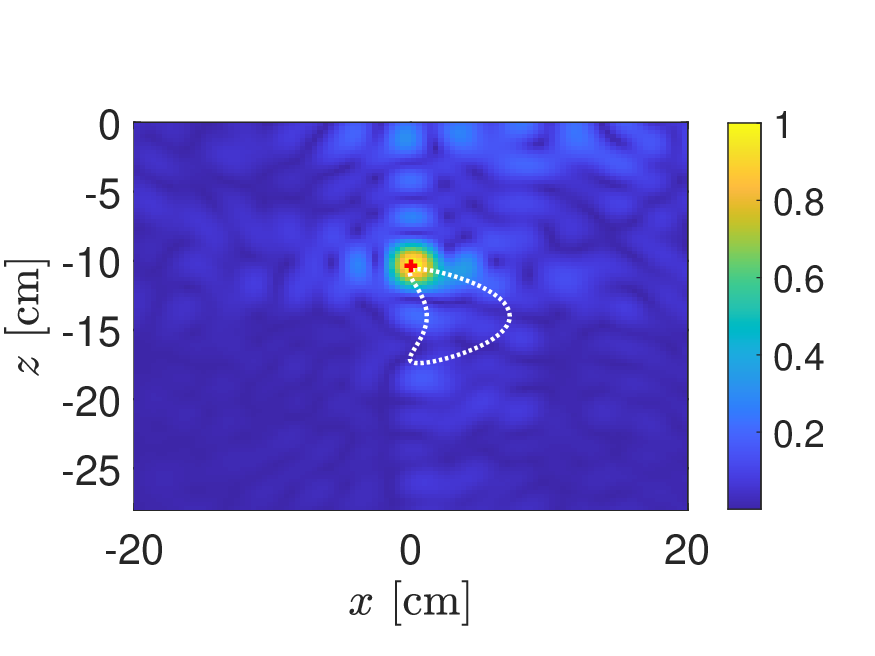}
    \caption{KM image}
  \end{subfigure}
  \caption{(a) Absolute difference ($\log_{10}$-scale) between the
    solution of the full boundary value problem and the
    first-order/flat interface approximation for scattered
    signals. (b) The resulting KM image using $\tilde{D}_{2}$ of the
    measurements using the first-order/flat interface approximation
    normalized by its maximum value. The location where the KM image
    attains its maximum is plotted as a red ``+'' symbol.}
  \label{fig:shower-curtain}
\end{figure}

In Fig.~\ref{fig:shower-curtain} we show results using this
first-order/flat interface approximation for the same problem as in
Fig.~\ref{fig:first-order}. The absolute difference
($\log_{10}$-scale) between the MFS solution for the full boundary
value problem and this first-order/flat interface approximation is
shown in Fig.~\ref{fig:shower-curtain}(a). In comparison with
Fig.~\ref{fig:first-order} we see that the error for this
approximation in the region above the air-soil interface is not
markedly different and suggests its validity for modeling
measurements.

To test its validity for modeling measurements for imaging, we show in
Fig.~\ref{fig:shower-curtain}(b) the resulting KM image normalized by
its maximum value using $\tilde{D}_{2}$ of the data simulated using
this approximation. The rough air-soil interface is the same one used
for the results shown in Fig.~\ref{fig:KM-firstorder}. Additionally,
we have added measurement noise to the approximate measurements so
that SNR $\approx 25$ dB just as with the results shown in
Fig.~\ref{fig:KM-firstorder}. This image formed using the
first-order/flat interface approximation to simulate measurements is
nearly indistinguishable with those shown in
Fig.~\ref{fig:KM-firstorder}. Moreover, the location where the KM
image attains its maxiumum value is exactly the same as that in
Fig.~\ref{fig:KM-firstorder}(a). In the context of faithfully
reproducing KM imaging results, these results show that this
first-order/flat interface approximation is sufficiently accurate.

\subsection{Point target}

All of the KM images we have shown exhibit a concentration about a
single location where the KM image attains maximum value. The image
for the rectangular target shown in Fig.~\ref{fig:KMshapes}(d) is the
most different in that the concentration of the image about that point
seems to follow the flat boundary. As we discussed in Section
\ref{sec:inverse-problem}, this GP-SAR problem is a severely limited
aperture imaging problem so we do not expect to recover much spatial
information about the subsurface target. In light of these severe
limitations, we consider simplifying a model of measurements by
restricting our attention to a point-like target.

We know from elementary scattering theory that the field scattered by
a particle in the far-field is proportional to the free-space Green's
function modified by a scattering amplitude~\cite{ishimaru}. It is the
scattering amplitude that contains information about the particle such as
its size, shape, and material properties. Limitations in GP-SAR
measurements result in a severely limited sample of the scattering
amplitude, but the KM images suggest the underyling Green's function
is robustly obtained.

Suppose we approximate the target below a rough air-soil interface by
a point target located at $\mathbf{r}_{0}^{\text{tar}}$ below a flat air-soil
interface.  To consider a point target model within the MFS
implementation we have used here, we solve \eqref{eq:approx-step1} as
before, but instead of solving \eqref{eq:approx-step2} and
\eqref{eq:approx-step3}, we instead solve
\begin{equation}
  \begin{bmatrix}
    -A_{1}^{\text{flat}} & A_{2}^{\text{flat}} \\
    -A_{3}^{\text{flat}} & A_{4}^{\text{flat}}  \\
  \end{bmatrix}
  \begin{bmatrix}
    \tilde{\mathbf{a}}^{(1)} \\ \tilde{\mathbf{b}}^{(1)}
  \end{bmatrix}
  = - \varrho u^{E}
  \begin{bmatrix}
    \mathbf{f}_{1} \\ \mathbf{f}_{2}
  \end{bmatrix},
  \label{eq:point-target}
\end{equation}
with
\begin{equation}
  u^{E} = \sum_{p = 1}^{P} G_{1}(\mathbf{r}^{\text{tar}}_{0} - (
  \mathbf{r}^{\text{int}}_{p} + \delta^{\text{int}} \hat{\mathbf{z}} ) )
  b_{p}^{(0)},
\end{equation}
denoting the field exciting the point target. The components of the
$P$-vectors $\mathbf{f}_{1}$ and $\mathbf{f}_{2}$ are given by
\begin{subequations}
  \begin{align}
    [ \mathbf{f}_{1} ]_{p}
    &= G_{1}( \mathbf{r}_{p}^{\text{int}} - \mathbf{r}_{0}^{\text{tar}})\\
    [ \mathbf{f}_{2} ]_{p}
    &= \partial_{n} G_{1}( \mathbf{r}_{p}^{\text{int}} - \mathbf{r}_{0}^{\text{tar}}),
  \end{align}
\end{subequations}
for $p = 1, \dots, P$, respectively. We call the scalar parameter
$\varrho$ the reflectivity which gives the scattering strength of the
point target.

\begin{figure}[htb]
  \centering
  \includegraphics[width=0.8\linewidth]{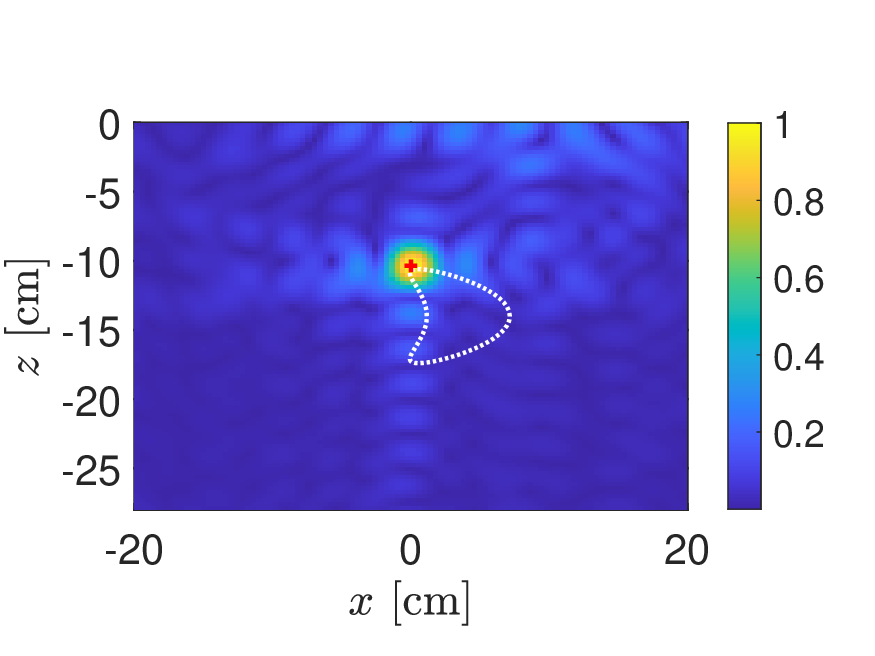}
  \caption{KM image for a point target located at
    $(x_{0}^{\text{tar}}, z_{0}^{\text{tar}}) = (0, -10.36)$ cm with reflectivity
    $\varrho = 8$ using $\tilde{D}_{2}$ of the measurements using the
    first-order/flat interface approximation with a point target. The
    KM image is normalized by its maximum value. The location where it
    attains its maximum value is plotted as a red ``+'' symbol.}
  \label{fig:point-target}
\end{figure}

In Fig.~\ref{fig:point-target} we show the KM image using the
first-order/flat interface approximation with a point target located
at $\mathbf{r}_{0}^{\text{tar}} = (0,-10.36)$ cm. This target location
corresponds to the location where the KM image attains its maximum
value using the full direct scattering problem measurements. The
reflectivity has been arbitrarily set to $\varrho = 8$. Additive noise
has been added to these approximate measurements so that SNR
$\approx 25$ dB.

The KM image shown in Fig.~\ref{fig:point-target} is nearly
indistinguishable with all other images shown here. This result
strongly indicates that this first-order/flat interface approximation
with a point target model captures the spatial information contained
in measurements based off of the full direct scattering
problem. Alternatively, these results suggest that any additional
sophistication in modeling measurements is lost when considering KM
images of targets.

\subsection{Modeling GP-SAR measurements}

Recall that entry $d_{mn}$ of the $M \times N$ data matrix $D$ gives
the measurements at frequency $\omega_{m}$ at spatial location
$\mathbf{r}_{n}$. We have discussed the following simplifying
approximations above.
\begin{enumerate}

\item[(i)] First-order target-interface interaction

\item[(ii)] Flat interface for scattering by the target

\item[(iii)] Point target model
  
\end{enumerate}
The results above show that within the context of KM imaging these
approximations are valid in that they faithfully capture the
qualitative features of KM images using data simulated from solutions
of the full boundary value problem. These approximations lead to the
following model for GP-SAR measurements of a subsurface target:
\begin{equation}
  D \approx D^{\text{model}} = R^{\text{rough}} + \varrho
  S^{\text{flat}}(\mathbf{r}_{0}),
  \label{eq:model}
\end{equation}
with the matrix $R^{\text{rough}}$ containing ground bounce signals by
the rough air-soil interface, and the matrix
$S^{\text{flat}}(\mathbf{r}_{0})$ containing  signals scattered by a
point target at location $\mathbf{r}_{0}$ with reflectivity
$\varrho = 1$ situated below a flat air-soil interface.

Within the context of this model, we can interpret the PCA and KM
imaging method we have discussed above as follows. After computing the
SVD of the measurements: $D = U \Sigma V^{H}$, we compute
\begin{equation}
  \tilde{D}_{J} = D - \sum_{j = 1}^{J} \sigma_{j} \mathbf{u}_{j}
  \mathbf{v}_{j}^{H} \approx \varrho S^{\text{flat}}(\mathbf{r}_{0}) +
  E,
\end{equation}
with
\begin{equation}
  E = R^{\text{rough}} - \sum_{j = 1}^{J} \sigma_{j}
  \mathbf{u}_{j} \mathbf{v}_{j}^{H},
  \label{eq:error}
\end{equation}
denoting the error in removing the ground bounce signal using the
contributions made by the first $J$ singular values.  When $E$ is
small, it acts effectively as additive noise to
$\varrho S(\mathbf{r}_{0})$. The evaluation of
$I^{\text{KM}}(\mathbf{y})$ given in \eqref{eq:KM-function} with
illuminations \eqref{eq:illuminations} is designed to ``unwind'' the
phase accumulated through propagation of the scattered field by a
point-like target below a flat air-soil interface at the search
location $\mathbf{y}$. Consequently, when the search point is in a
neighborhood of the point target position, the phases of the data and
illuminations match and lead to the absolute value $|I^{\text{KM}}|$
developing a peak.

By systematically testing and validating each of the simplifying
assumptions above, we have established the validity of model
\eqref{eq:model}. A key point to this model is the recognition of the
limitations in GP-SAR measurements and understanding how those
limitations affect the overall information content in those
measurements. This model provides valuable insight into how the PCA
and KM imaging method we have used here works.

\subsection{Extensions}

Although our study uses only simulations of the direct scattering
problem for scalar waves in two dimensions, the principles behind
these basic assumptions extend naturally to more complicated
three-dimensional scattering problems. For that case, the entries of
matrices $R$ and $S(\mathbf{r}_{0})$ correspond to the respective
field values for the three-dimensional problem. Moreover, measurement
model \eqref{eq:model} can be modified to incorporate more
sophistication when that is necessary. For example, we can include
additional $S$ matrices for multiple subsurface targets.  If the point
target model is not valid for a related, but different problem, one
can replace the matrix $S$ with a more sophisticated scattering
model.

We have recently proposed a dispersive point target model where the
reflectivity $\varrho$ varies with
frequency~\cite{KT:dispersive}. That target model is easily
incorporated in \eqref{eq:model} as
\begin{equation}
  D^{\text{model}} = R^{\text{rough}} + \text{diag}(\varrho_{1},
  \dots, \varrho_{M}) S^{\text{flat}}(\mathbf{r}_{0}),
\end{equation}
where $\varrho_{m}$ corresponds to the reflectivity at frequency
$\omega_{m}$. From this dispersive point target model, we easily
extend it further to include anisotropy in the reflectivity using
\begin{equation}
  D^{\text{model}} = R^{\text{rough}} + F \odot
  S^{\text{flat}}(\mathbf{r}_{0}),
  \label{eq:full-model}
\end{equation}
with $\odot$ denoting the Hadamard (element-wise) matrix product. The
entries of the $M \times N$ matrix $F$ are the reflectivities
$\varrho_{mn}$ which now vary in frequency and space.  These two
extensions open opportunities for quantitative imaging methods that
seek to determine reflectivities rather than seeking to determine
spatial information about targets. Given the inherent space/angle
limitations in GP-SAR measurements, these quantitative imaging
problems may be more practically useful than seeking to find more
spatial information about targets.

\begin{figure}[htb]
  \centering
  \begin{subfigure}[t]{0.48\linewidth}
    \includegraphics[width=\linewidth]{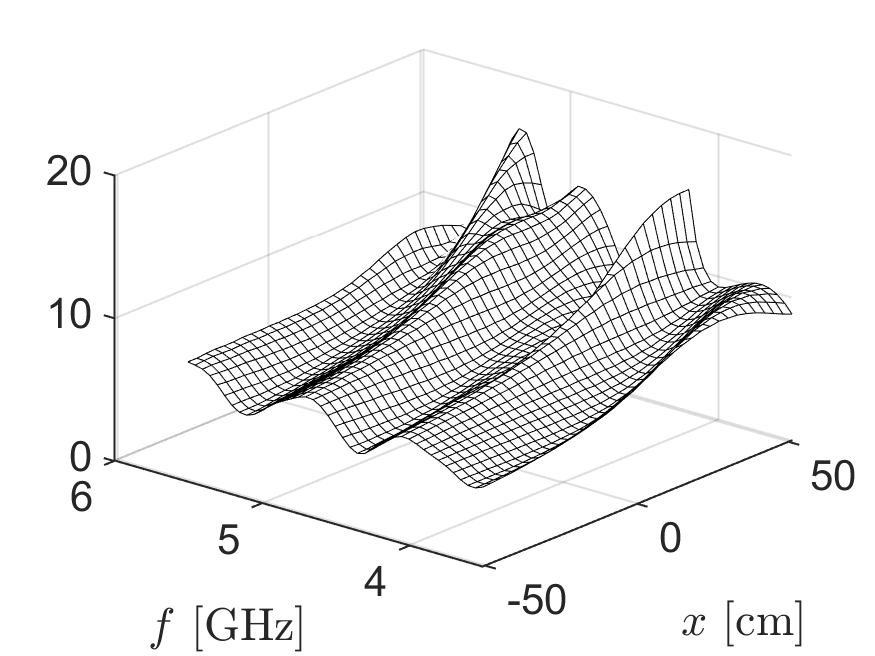}
    \caption{model}
  \end{subfigure}
  \begin{subfigure}[t]{0.48\linewidth}
    \includegraphics[width=\linewidth]{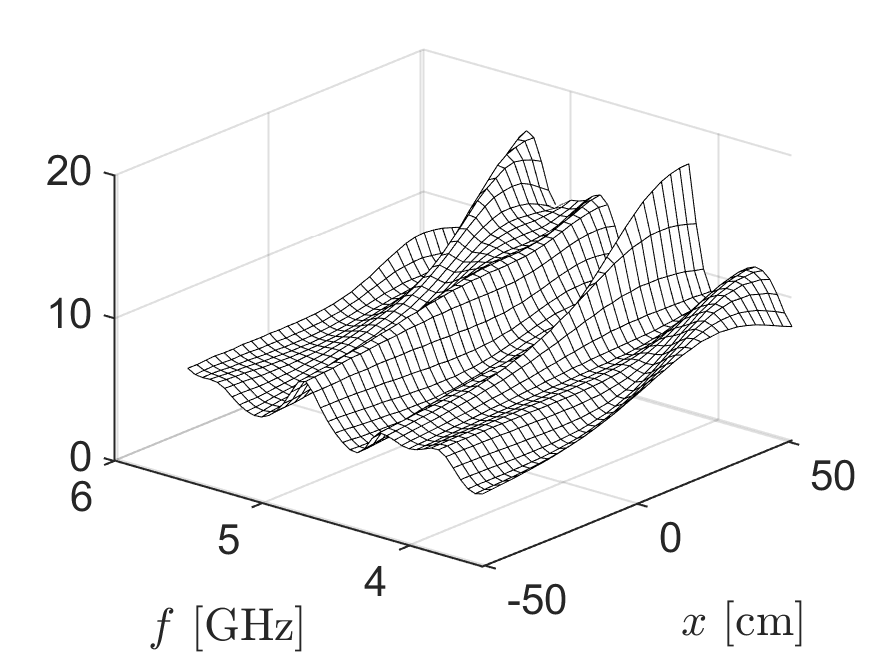}
    \caption{data}
  \end{subfigure}
  \caption{Plots of the absolute values of the entries of the
    reflectivity matrix $F$, as a function of frequency $f$ and
    spatial location $x$ of measurements. The left plot (``model'')
    shows results computed using the first-order target-interface and
    target scattering below a flat air-soil interface assumptions. The
    right plot (``data'') shows results computed using measurements
    computed from the full direct scattering problem.}
  \label{fig:reflectivity-matrix}
\end{figure}

To investigate the validity of \eqref{eq:full-model}, we consider
$R^{\text{rough}}$ and $S^{\text{flat}}(\mathbf{r}_{0}^{\text{tar}})$
that we used to evaluate the point target assumption in
Fig.~\ref{fig:point-target}. We then consider the measurements used
for Fig.~\ref{fig:shower-curtain} that used the first-order
target-interface approximation and scattering by the kite-shaped
target below a flat air-soil interface. To determine the reflectivity
matrix $F$, we subtract $R^{\text{rough}}$ from those measurements and
then divide that result element-wise by the entries of
$S^{\text{flat}}(\mathbf{r}_{0}^{\text{tar}})$. A plot of the absolute
values of this result are shown in
Fig.~\ref{fig:reflectivity-matrix}(a), which we have labelled as
``model.'' We do the same computation using the measurements from the
full direct scattering problem and those results are shown in
Fig.~\ref{fig:reflectivity-matrix}(b), which we have labelled as
``data.'' These results show that the measurements computed from full
direct scattering problem can be modeled accurately using
\eqref{eq:full-model} which includes the reflectivity matrix $F$.

\begin{figure}[htb]
  \centering
  \includegraphics[width=0.48\linewidth]{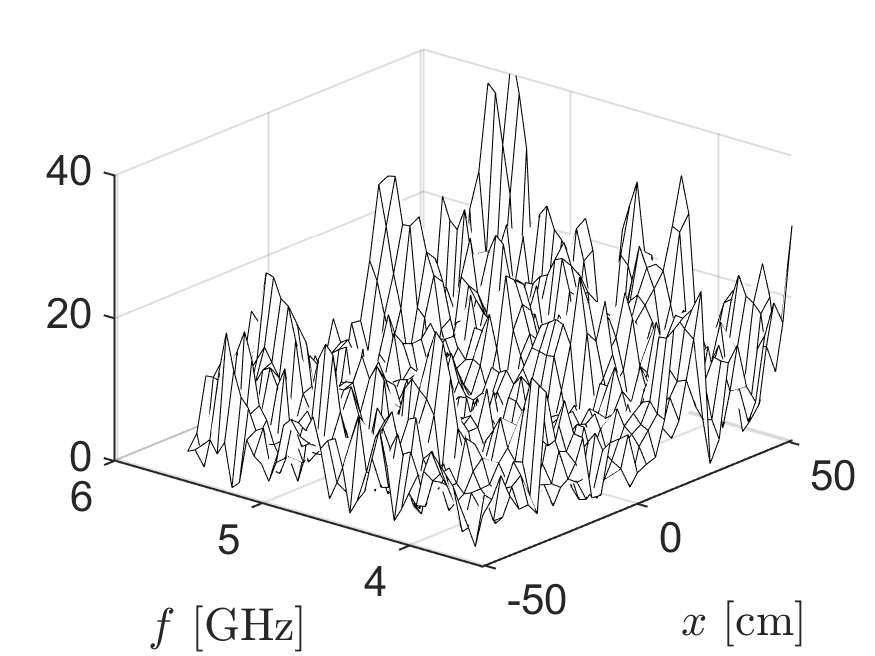}
  \includegraphics[width=0.48\linewidth]{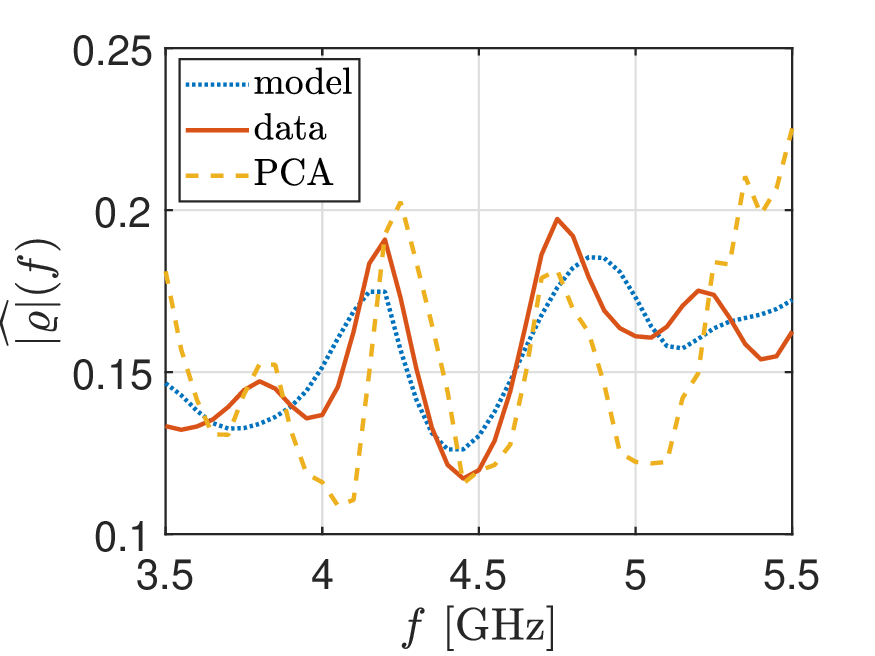}
  \caption{[Left] Estimated absolute values of the entries of the
    reflectivity matrix after applying PCA. [Right] A comparison of
    the normalized spatial averages of the absolute values of the
    reflectivity matrices, denoted by $\widehat{|\varrho|}(f)$ for the
    two results shown in Fig.~\ref{fig:reflectivity-matrix} and the results
    shown here.}
  \label{fig:PCA-reflectivity}
\end{figure}

The results shown in Fig.~\ref{fig:reflectivity-matrix} make explicit
use of $R^{\text{rough}}$ which tacitly implies that the rough
air-soil interface is known. We now assume it is not known and apply
PCA as we have done above by truncating the contributions made by the
first two singular values to obtain $\tilde{D}_{2}$. For that case, we
estimate the reflectivity matrix $F$ by dividing $\tilde{D}_{2}$
element-wise by the entries of
$S^{\text{flat}}(\mathbf{r}_{0}^{\text{tar}})$. A plot of the absolute
values of this result are shown in the right plot of
Fig.~\ref{fig:PCA-reflectivity}. This result shows that this estimate
has large, spurious oscillations due to the error $E$ introduced in
\eqref{eq:error}. Although $E$ only acts as additive measurement noise
for imaging using KM, its effect on estimating $F$ here is much
stronger. In light of these strong oscillatory errors, we consider the
spatial average of absolute values:
\begin{equation}
  \overline{| \varrho |}(f_{m}) = \frac{1}{N} \sum_{n = 1}^{N} |
  \varrho_{mn} |.
\end{equation}
Then we normalize this spatial average by its $2$-norm and denote it
by $\widehat{| \varrho |}(f)$. A plot of $\widehat{| \varrho |}(f)$
for the results shown in Fig.~\ref{fig:reflectivity-matrix} and this
PCA result are shown in the right plot of
Fig.~\ref{fig:PCA-reflectivity}. Even though these normalized
frequency spectra do not quantitatively agree with one another, they
share several characteristic features, especially within the sub-band
between $4$ GHz and $5$ GHz.

\begin{figure}[htb]
  \centering
  \includegraphics[width=0.8\linewidth]{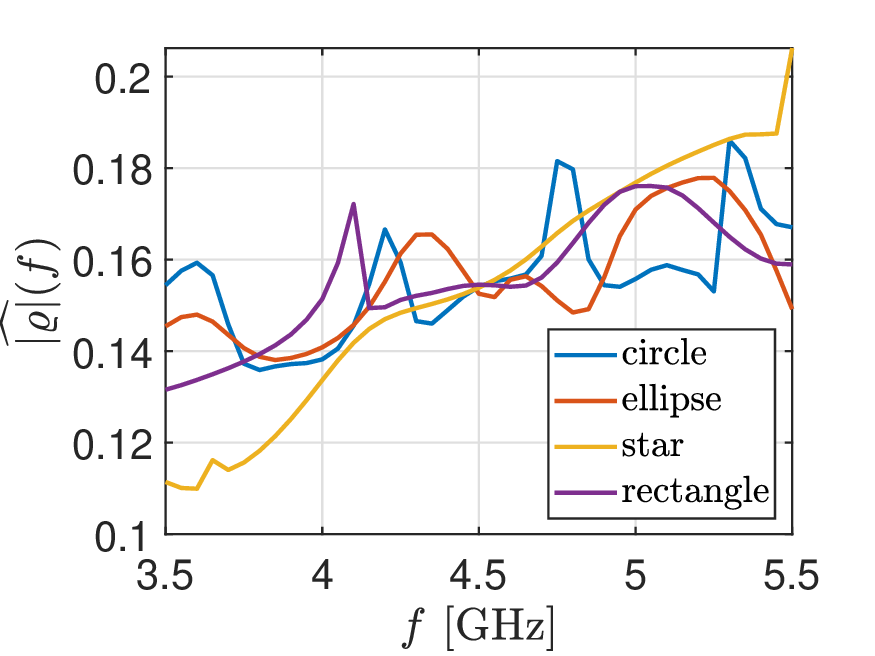}
  \caption{Frequency spectra $\widehat{|\varrho|}(f)$ for the
    different shaped targets shown in Fig.~\ref{fig:KMshapes}.}
  \label{fig:spectra}
\end{figure}

In Fig.~\ref{fig:KMshapes} we showed that the imaging method we use
here does not produce images that show clear differences for markedly
different target shapes. In Fig.~\ref{fig:spectra} we show
$\widehat{|\varrho|}(f)$ for each of these targets. These results show
that these frequency spectra are markedly different for the different
targets.  We have recently studied how these normalized frequency
spectra can be used to classify subsurface
targets~\cite{KT:classification}. From these results we see that there
is characterizing information contained in GP-SAR measurements. By
considering \eqref{eq:full-model}, we are able to propose novel
quantitative methods that enable target classification.

\section{Conclusions}
\label{sec:conclusions}

We have studied GP-SAR imaging of subsurface targets below a rough
air-soil interface. The imaging method we have used first uses
principal component analysis to approximately remove ground bounce
signals. Then it uses Kirchhoff migration on this processed data to
identify and locate targets.  By systematically studying and
validating simplifying assumptions for modeling measurements, we have
proposed the simplified model \eqref{eq:model}.  This model is the sum
of ground bounce signals by the rough air-soil interface and scattered
signals by a point target below a flat air-soil interface.  This
simple model provides valuable insight into how the imaging method we
have used here works.

The first assumption we discussed uses only first-order interactions
between the target and the air-soil interface. Using first-order
interactions is often done in inverse scattering problems for waves in
a multiple scattering medium. The key to this approximation is that
first-order interactions still retain crucially important phase
information from scattering by the target. This phase information
corresponds to discontinuities at boundaries and interfaces enabling
the identification and location of targets.

The assumption that measurements of the signals scattered by the
subsurface target can be modeled using a flat air-soil interface is a
consequence of the so-called shower curtain effect. Because
measurements are typically taken farther away from the air-soil
interface than the distance between the target and interface, the
effect of surface roughness on measurements is relatively small. This
assumption is important because it relieves an imaging method from
having to know the roughness of the air-soil interface.

The most extreme assumption used in this model is the point target
model. The point target assumption is valid because of the inherent
limitations in synthetic aperture measurements. Because this imaging
problem is a severely limited aperture imaging problem, we do not
expect to recover a lot of spatial information about the target. We
have shown for the imaging method we have used here that we can
recover only a representative point for the target. Consequently, a
more sophisticated target model is not necessary.

Beyond the specific details of measurement model \eqref{eq:model}, it
gives a theoretical framework for studying subsurface imaging
problems. Moreover, the extended model \eqref{eq:full-model} opens up
new opportunities for quantitative imaging methods that may enable
target classification.  For these reasons, we believe that measurement
model \eqref{eq:model} is useful for studying GP-SAR imaging of
subsurface targets.

\bibliographystyle{IEEEtran}
\bibliography{KT-2024}

\end{document}